\documentclass[letterpaper, aps, tightenlines, nofootinbib, 11pt]{article}
\pdfoutput = 1

\usepackage{setspace}
\usepackage[bookmarks = false, colorlinks = true, linkcolor = blue, citecolor = purple]{hyperref}
\usepackage{shortcuts}
\usepackage{titlesec}
\usepackage{cite}

\usepackage[margin = 2.5cm]{geometry}
\def\thesection{\arabic{section}}
\def\thesubsection{\arabic{subsection}}
\titleformat{\section}
  {\normalfont\Large\bfseries}{\thesection}{1em}{}
\titleformat{\subsection}
  {\normalfont\large\it}{\thesection.\thesubsection}{1em}{}
\titleformat{\subsubsection}
  {\normalfont\it}{\thesubsection.\thesubsubsection}{1em}{}
\numberwithin{equation}{section}

\begin{document}

\begin{center}

\thispagestyle{empty}

\vspace*{5em}

{ \LARGE {\bf Chern-Simons theory coupled to bifundamental scalars}}

\vspace{1cm}

{\large Shamik Banerjee and \DJ or\dj e Radi\v cevi\'c}
\vspace{1em}

{\it Stanford Institute for Theoretical Physics and Department of Physics\\ Stanford University \\
Stanford, CA 94305-4060, USA}\\
\vspace{1em}
\texttt{bshamik@stanford.edu,\ djordje@stanford.edu}\\
 \end{center}

\vspace{0.08\textheight}
\begin{abstract}
    We study the three-dimensional theory of two Chern-Simons gauge fields coupled to a scalar field in the bifundamental representation of the $SU(N)_k \times SU(M)_{-k}$ gauge group. At small but fixed $M \ll N$, this system approaches the theory of a Chern-Simons field coupled to fundamental matter, conjectured to be dual to a parity-violating version of Vasiliev's higher-spin gauge theory in AdS$_4$. At finite $M/N$ and large 't Hooft coupling this theory (or its SUSY version) is expected to be dual to an Einstein-like gravity.  We show at two loops that this theory possesses  a line of fixed points at any value of $M/N$. We also prove that turning on a finite but small $M/N$ gaps out the light states that Chern-Simons theory coupled to fundamental matter develops when placed on a torus. We also comment on the higher genus case.
\end{abstract}

\pagebreak
\setcounter{page}{1}

\section{Introduction}

Recent years have seen much progress in understanding the holographic duality between 3D vector-like theories and Vasiliev's higher-spin gauge theories in 4D anti-de Siter space (AdS$_4$) \cite{Vasiliev:1992av, Vasiliev:1995dn, Vasiliev:1999ba, Vasiliev:2003ev, Sezgin:2002ru, Giombi:2012ms}.\footnote{There have also been very interesting parallel developments of this story. The dual of the three-dimensional higher spin gauge theory on $AdS_3$ has been identified as the large-$N$ limit of $W_N$ minimal models in two dimensions \cite{Gaberdiel:2010pz}, and these theories have also been related to a string theory on $AdS_3$ \cite{Gaberdiel:2013vva} (see also \cite{Chang:2011mz, Chang:2013izp}). The status and applicability of the higher-spin realization of dS/CFT correspondence has also been fleshed out \cite{Anninos:2011ui, Ng:2012xp, Anninos:2012ft, Anninos:2013rza, Banerjee:2013mca, Das:2012dt}.} According to the Klebanov-Polyakov conjecture, the singlet operators of the 3D $O(N)$ or $U(N)$ vector model at large $N$ are dual to the Vasiliev higher spin fields in the bulk AdS$_4$ \cite{Klebanov:2002ja, Sezgin:2002rt, Petkou:2003zz, Sezgin:2003pt,  Girardello:2002pp, Giombi:2009wh, Giombi:2010vg, Giombi:2011ya, Chang:2012kt}.\footnote{ Refs.~\cite{Das:2003vw, Koch:2010cy, Douglas:2010rc, Jevicki:2011ss} have attempted to derive this duality from first principles.} The singlet constraint can be dynamically imposed by gauging the global symmetry and adding a Chern-Simons (CS) kinetic term for the gauge field \cite{Giombi:2009wh, Giombi:2011kc, Aharony:2011jz, Shenker:2011zf}. The singlet (CS-gauged) vector models constructed in this way are conjectured to be dual to the parity-violating Vasiliev theory in AdS$_4$.

Singlet vector models can also be studied on higher genus spatial surfaces \cite{Banerjee:2012gh}. In particular, on a spatial torus, these theories contain a set of states whose energies scale as $1/N$. These states are exactly degenerate at $N=\infty$, which corresponds to a classical theory in the bulk. Such light states arise due to the non-trivial dynamics of the CS theory on spaces with a non-trivial fundamental group \cite{Witten:1988hf, Elitzur:1989nr}; they are not present in the Vasiliev theories studied before, and remain mysterious from the bulk point of view. Any bulk theory that is conjectured to be dual to the singlet vector model on the boundary should be able to accommodate these states.

Ref.~\cite{Chang:2012kt} has recently made a very interesting proposal of the duality between supersymmetric versions of Vasiliev theory in the bulk and bifundamental CS-matter theory on the boundary with a simply connected spatial manifold. In this paper we supplement their results by studying a simple, non-supersymmetric bifundamental CS-matter model in various spacetimes. We focus on a theory with the gauge group $SU(N)_k\times SU(M)_{-k}$, where $(k,-k)$ are the levels of the $(SU(N),SU(M))$ CSterms. The matter is a scalar field transforming in the bifundamental representation $(N,\bar M)$ of the gauge group, and we work in the 't Hooft limit where we take $N$, $M$, and $k$ to infinity while keeping $\lambda = N/k$ and $\xi = M/N$ fixed. At $N = \infty$, the two regimes of interest are $\xi = 0$, corresponding to a finite $M$, or $\xi > 0$, corresponding to the double-scaling limit described above. In the bulk, these two regimes map to regimes of zero and non-zero bulk 't Hooft coupling, respectively.\footnote{This is so because, in the bulk, the $SU(M)$ group can be understood as a gauge group with gauge coupling $1/N$. Please see \cite {Chang:2012kt} for further details.}  In the limit when $\xi$ is small the difference from the fundamental vector model with gauge group $SU(N)_{k}$ is expected to be small. To see finite-$M$ effects as $\xi$ is taken to zero, one must keep track of $O(1/N)$ corrections in the bifundamental computation, which we do not do in this paper.

Various aspects of CS theory coupled to fundamental matter have been studied in detail in \cite{Giombi:2011kc, Aharony:2011jz, Shenker:2011zf, Maldacena:2011jn, Maldacena:2012sf, Aharony:2012nh, Banerjee:2012gh, Radicevic:2012in, Chen:1992ee, Aharony:2012ns, Jain:2013py, Jain:2013gza, Takimi:2013zca, Jain:2012qi, Yokoyama:2012fa, GurAri:2012is, Giombi:2013yva}. In Section \ref{sec:feyn2loops} we study, following \cite{Chang:2012kt, Aharony:2011jz}, the non-supersymmetric CS theory coupled to bifundamental scalar matter. We compute the two-loop $\beta$-functions of the theory and find two lines of fixed points parametrized by the gauge coupling $\lambda$. In Section \ref{sec:torus}, following \cite{Banerjee:2012gh}, we study the CS-bifundamental theories on a spatial torus. Encouragingly, we find that there are no exactly degenerate states in the 't Hooft limit. Instead, we find that the gap is proportional to $\xi$, when $\xi$ is small, and along the way we develop a straightforward diagrammatic way of arranging the perturbation
theory in $\xi$. Finally, in the concluding section, we point out how these results suggest that bifundamental theories can be used to regulate fundamental matter theories by tuning the ratio $\xi$ of the two ranks in a bifundamental theory. On a spatial torus, by changing $\xi$ one tunes the mass of the gauge field holonomies and, at $\xi \sim 1/N$, transitions into a phase that looks like the fundamental theory (i.e.~the singlet vector model) on a torus. This indicates that the heretofore mysterious modification of Vasiliev theory that can accommodate for the light states found in \cite{Banerjee:2012gh} should be attainable as a particular limit of the bulk theory studied in \cite{Chang:2012kt}.

A note on conventions is in order. The two papers on which our analysis rests, \cite{Aharony:2011jz} and \cite{Banerjee:2012gh}, employ different conventions and have slightly different actions. In order to be able to check that our computations correctly reduce to the results of these two papers, we work with different actions in different sections of this paper; in each section we follow the conventions of its guiding reference.

\section{Perturbative fixed points} \label{sec:feyn2loops}

We consider an $SU(N)_k \times SU(M)_{-k}$ Euclidean CS theory coupled to scalars in the bifundamental representation $(N,\bar M)$. Without loss of generality we will take $M \leq N$. The partition function of our theory is
\bel{
  Z = \int[\d A\, \d B\, \d \phi] e^{-S},
}
and the action that contains all the marginal terms is
\algnl{\notag
  S
  &= \frac{ik}{8\pi} \int \Tr_N \left(A\d A + \frac{2i}3 A^3\right) - \frac{ik}{8\pi} \int \Tr_M \left(B\d B + \frac{2i}3 B^3\right) + \int \d^3 x\  \Tr_M \left((D_\mu \phi)\+ D^\mu \phi \right) + \\ \label{def S}
  &\quad + \int \d^3 x \left\{\frac{g_1}{3}\, \Tr_M\left(\phi\+ \phi\right)^3 + \frac{g_2}{2}\,\Tr_M\left(\phi\+ \phi\right)\Tr_M\left(\phi\+ \phi\right)^2 + \frac{g_3}{6} \left[\Tr_M\left(\phi\+ \phi\right)\right]^3\right\}.
}
The covariant derivative is
\bel{
  D_\mu \phi = \del_\mu \phi - i A_\mu \phi + i \phi B_\mu.
}
The bifundamental scalar, $\phi$, is an $N\times M$ matrix transforming as $\phi \mapsto U \phi V\+$ for $U \in SU(N)$, $V \in SU(M)$, and $A$ and $B$ are Hermitian connections transforming as $A \mapsto U A U\+ + i U \d U\+$ and $B \mapsto V B V\+ + i V \d V\+$. The traces $\Tr_{N/M}$ are taken in the fundamental representations of $SU(N)$ and $SU(M)$, respectively. Lie algebra conventions and the Feynman rules stemming from this action are collected in
Appendix \ref{sec:FeynRules}.

In this section we perform the fixed point analysis of the 't Hooft limit perturbatively in $\lambda = N/k$. We work to two-loop order, closely following the work of \cite{Aharony:2011jz}. The CS coupling $g^2 = 4\pi/k$ is quantized and does not run under RG flows; hence we only need to compute the $\beta$-functions for $g_1$, $g_2$, and $g_3$. These are found by computing the two-loop amplitude
\bel{\label{def M}
  \M = \avg{\phi_{ii'}\+ \phi_{jj'} \phi_{kk'}\+ \phi_{ll'} \phi_{mm'}\+ \phi_{nn'}}
}
when all external momenta are zero.  At tree level, this amplitude is given by,
\algnl{\notag
  \M\_{tree} =
  &- g_1 \left(\delta_{i'j'} \delta_{jk} \delta_{k'l'} \delta_{lm} \delta_{m'n'} \delta_{ni} + \trm{11\  permutations}\right) - \\ \notag
  &- g_2 \left( \delta_{i'j'} \delta_{jk} \delta_{k'l'} \delta_{li} \delta_{m'n'} \delta_{nm} + \trm{17\ permutations}\right) - \\
  &- g_3\left( \delta_{i'j'} \delta_{ji} \delta_{k'l'} \delta_{lk} \delta_{m'n'} \delta_{nm} + \trm{5\ permutations} \right).
}
Once loop corrections are introduced, the amplitude must be regulated by counterterms $\delta g_i(\mu)$ multiplying the above tensor structures. These counterterms depend on the subtraction scale $\mu$ and can be used to extract the $\beta$-functions $\beta_i$ of all three couplings. We will use dimensional reduction\footnote{The CS term cannot be written in general $d\ne 3$ dimensions, and so the standard dimensional regularization cannot be applied to CS-matter theories. Instead, one uses the dimensional reduction scheme where the tensor algebra appearing in the Feynman integrals is done in three dimensions and then the resulting scalar integral is analytically continued to general space dimensions. This preserves gauge invariance at least up to two loops. See \cite{Chen:1992ee} for details.} to regulate the UV divergences, and we will work in the minimal subtraction scheme.

\begin{figure}
  \centering
  \includegraphics[width = \textwidth, keepaspectratio = true]{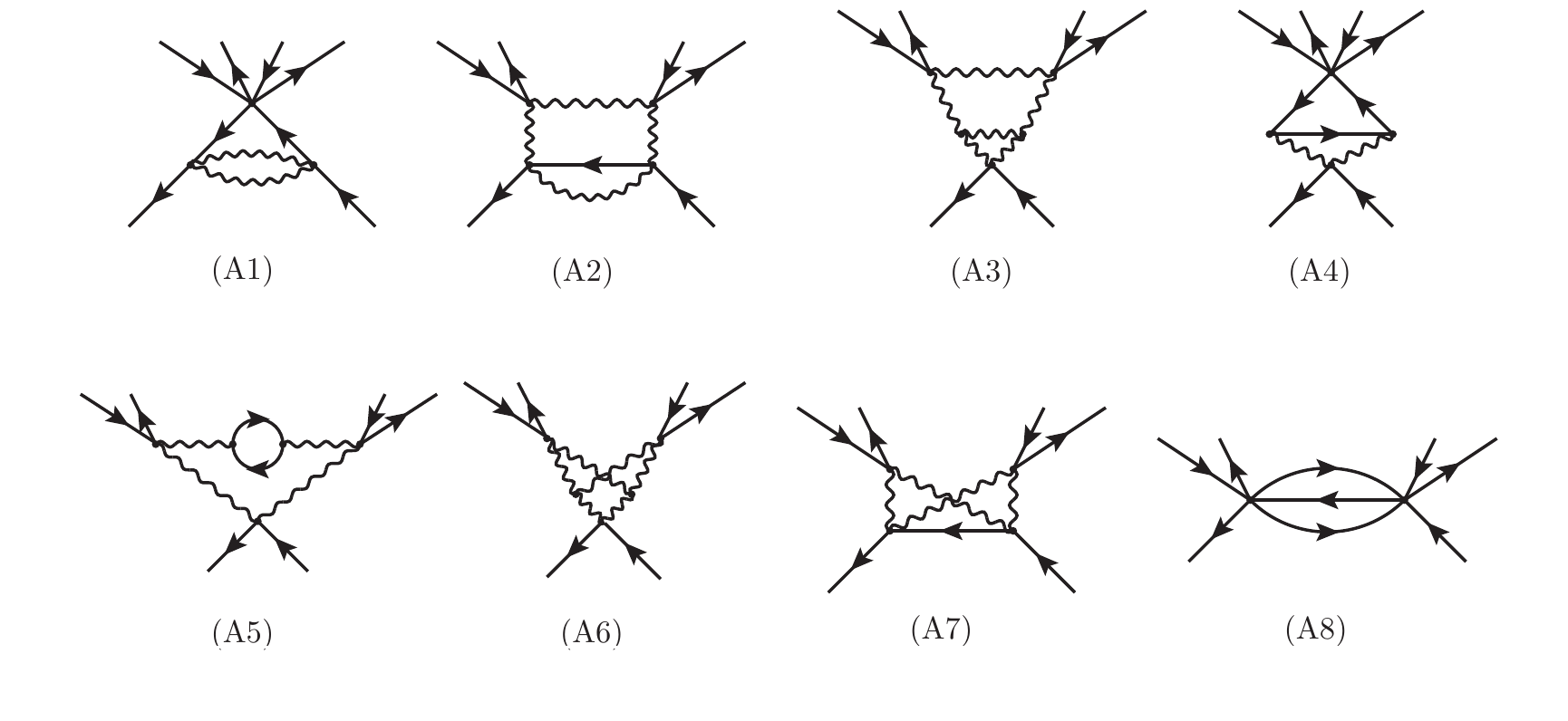}\\
  \caption{\small The Feynman diagrams that contribute to the flowing of six-point couplings at two loops. This representation is schematic. Each six-point coupling in diagrams (A1), (A4), and (A8) must be treated as a single-, double-, or triple-trace coupling in turn. Similarly, each gauge boson must be treated as either an $A_\mu$ or a $B_\mu$ boson.}\label{fig:feyn2loop}
\end{figure}

Despite the apparent complexity of this computation, Aharony et al.~have shown that, remarkably, only eight diagrams need to be computed when dealing with fundamental CS-matter to two-loop order \cite{Aharony:2011jz}.\footnote{In an earlier version of this paper, a diagram that is zero for $O(N)$ was mistakenly concluded to be zero for $U(N)$, based on the findings in \cite{Aharony:2011jz}. We thank Guy Gur-Ari and Raghu Mahajan for extensive discussions that have clarified this issue.} The same argument goes through for bifundamental CS-matter (Fig.~\ref{fig:feyn2loop}). The additional complication in our case is the existence of two gauge fields and two additional multi-trace scalar six-point couplings, so each diagram in the fundamental matter theory can be thought to generate a number of related diagrams in the bifundamental theory, each with the same momentum structure but with different index contractions and multiplicity. These are all straightforward to enumerate and compute. The Feynman rules are given in Appendix \ref{sec:FeynRules}, and applying them gives
%\algnl{\notag
 % \trm{(A1)}: \quad
  %& \delta g_1 = -\frac34 g_1 g^4 (N^2 + M^2)I_{A1},\ \ \delta g_2 = g^4 \left[\frac12 g_1 (N + M) - \frac 34 g_2 (N^2 + M^2 - 4NM) \right] I_{A1},\\
  %&\delta g_3 = g^4 \left[6g_1 + 3(N + M) g_2 - \frac34 g_3 (N^2 + M^2 - 4NM) \right] I_{A1},\\
  %\trm{(A2)}: \quad
  %& \delta g_1 = \frac3{16} g^8 \left(N^2 + M^2 - 8NM\right) I_{A2},\ \ \delta g_2 = \frac{29}8 g^8 (N + M) I_{A2},\ \ \delta g_3 = -\frac34 g^8 I_{A2}.\\
  %\trm{(A3)}: \quad
  %& \delta g_1 = -\frac3{16} g^8 \left(N^2 + M^2\right) I_{A3},\ \ \delta g_2 = - \frac18 g^8 (N + M) I_{A3},\ \ \delta g_3 = \frac34 g^8 I_{A3},\\
  %\trm{(A4)}: \quad
  %& \delta g_1 = 6g_1g^4 NM I_{A4},\ \ \delta g_2 = 6  g^4\left(g_1 (N + M) + g_2 NM \right) I_{A4},\ \ \delta g_3 = 6g_3 g^4 NM I_{A4},\\
  %\trm{(A5)}: \quad
  %& \delta g_1 = -3 g^8 NM I_{A5},\ \ \delta g_2 = g^8 (N + M) I_{A5}, \ \ \delta g_3 = 0,\\ \
  %\trm{(A6)}: \quad
  %& \delta g_1 = 0,\ \ \delta g_2 = -\frac54 g^8 (N + M) I_{A6},\ \ \delta g_3 = -\frac{21}2 g^8 I_{A6},\\
  %\trm{(A7)}: \quad \notag
  %& \delta g_1 = 3 g_1^2 NM I_{A7},\ \delta g_2 = \left(3g_1^2(N + M) + 4g_1g_2 NM \right) I_{A7},\\
  %&\delta g_3 = 6 \left(g_1^2 + g_1 g_2 (N + M) + g_2^2 NM \right) I_{A7}.
%}
\algnl{\notag
  \trm{(A1)}: \quad
  & \delta g_1 = - 3 g_1 g^4 (N^2 + M^2)I_{A1},\ \ \delta g_2 = g^4 \left[2 g_1 (N + M) - 3 g_2 (N^2 + M^2 - 4NM) \right] I_{A1},\\
  &\delta g_3 = g^4 \left[24 g_1 + 12 (N + M) g_2 - 3 g_3 (N^2 + M^2 - 4NM) \right] I_{A1},\\
  \trm{(A2)}: \quad
  & \delta g_1 = 3 g^8 \left(N^2 + M^2 - 8NM\right) I_{A2},\ \ \delta g_2 = 58 g^8 (N + M) I_{A2},\ \ \delta g_3 = -12 g^8 I_{A2}.\\
  \trm{(A3)}: \quad
  & \delta g_1 = -3 g^8 \left(N^2 + M^2\right) I_{A3},\ \ \delta g_2 = - 2 g^8 (N + M) I_{A3},\ \ \delta g_3 = 12 g^8 I_{A3},\\ \notag
  \trm{(A4)}: \quad
  & \delta g_1 = 24 g_1g^4 NM I_{A4},\ \ \delta g_2 = 24  g^4\left(g_1 (N + M) + g_2 NM \right) I_{A4},\\
  & \delta g_3 = 24 g_3 g^4 NM I_{A4},\\
  \trm{(A5)}: \quad
  & \delta g_1 = - 48 g^8 NM I_{A5},\ \ \delta g_2 = 16 g^8 (N + M) I_{A5}, \ \ \delta g_3 = 0,\\
  \trm{(A6)}: \quad
  &\delta g_1 = 0,\ \ \delta g_2 = -2(N + M) g^8 I_{A6},\ \ \delta g_3 = -20 g^8 I_{A6}, \\
  \trm{(A7)}: \quad
  & \delta g_1 = 0,\ \ \delta g_2 = - 20 g^8 (N + M) I_{A7},\ \ \delta g_3 = - 168 g^8 I_{A7},\\
  \trm{(A8)}: \quad \notag
  & \delta g_1 = 3 g_1^2 NM I_{A8},\ \delta g_2 = \left(3g_1^2(N + M) + 4g_1g_2 NM \right) I_{A8},\\
  &\delta g_3 = 6 \left(g_1^2 + g_1 g_2 (N + M) + g_2^2 NM \right) I_{A8}.
}

The quantities $I_{Ai}$ label the dimensionally regulated momentum space integrals, which evaluate to
\bel{
  I_{A1} = I_{A2} = I_{A3} = I_{A5} = I_{A8} = \frac1{32\pi^2} \frac1\eps,\ I_{A4} = \frac1{16\pi^2}\frac1\eps,\ I_{A6} = -\frac3{64\pi^2} \frac1\eps,\ I_{A7} = \frac3{64\pi^2} \frac1\eps.
}
Now we can switch to the 't Hooft couplings,
\bel{
  \lambda = g^2 N,\quad \lambda_1 = g_1N^2,\quad \lambda_2 = g_2N^2 M,\quad \lambda_3 = g_3N^2M^2,
}
and find that the counterterms are given by
%\algnl{\notag
 % \delta \lambda_1 &= \frac3{128\pi^2 \eps} \Big[4\,\xi \lambda_1^2 - \left(1 - 16\xi + \xi^2 \right) \lambda_1 \lambda^2 - 6\,\xi \lambda^4 \Big], \\ \label{delta lambda}
  %\delta \lambda_2 &= \frac1{256 \pi^2 \eps} \Big[32\, \xi \lambda_1 \lambda_2 + 100\, \xi(1 + \xi) \lambda_1 \lambda^2 - 6 \left(1  - 20\xi + \xi^2 \right) \lambda_2 \lambda^2 + 21\,\xi(1 + \xi) \lambda^4  \Big], \\ \notag
  %\delta \lambda_3 &= \frac3{128\pi^2\eps} \Big[8\,\xi^2 \lambda_1^2 + 8\,\xi (1 + \xi) \lambda_1 \lambda_2 + 8\,\xi \lambda_2^2\ + \\ \notag
 % &\qquad \qquad \qquad + 8\, \xi^2 \lambda_1 \lambda^2 + 4\, \xi(1 + \xi) \lambda_2 \lambda^2 -  \left(1 - 20\xi + \xi^2\right) \lambda_3 \lambda^2 - 21\, \xi^2 \lambda^4 \Big].
%}
\algnl{\notag
  \delta \lambda_1 &= \frac3{32\pi^2 \eps} \Big[\xi \lambda_1^2 - \left(1 - 16\xi + \xi^2 \right) \lambda_1 \lambda^2 - 24\,\xi \lambda^4 \Big], \\ \notag
  \delta \lambda_2 &= \frac1{32 \pi^2 \eps} \Big[3\, \xi(1 + \xi) \lambda_1^2 + 4\, \xi \lambda_1 \lambda_2 + 50\, \xi(1 + \xi) \lambda_1 \lambda^2 - 3 \left(1  - 24\xi + \xi^2 \right) \lambda_2 \lambda^2 + 33\,\xi(1 + \xi) \lambda^4  \Big], \\ \notag
  \delta \lambda_3 &= \frac3{32\pi^2\eps} \Big[2\,\xi^2 \lambda_1^2 + 2\,\xi (1 + \xi) \lambda_1 \lambda_2 + 2\,\xi \lambda_2^2\ + \\  \label{delta lambda}
  &\qquad \qquad \qquad + 8\, \xi^2 \lambda_1 \lambda^2 + 4\, \xi(1 + \xi) \lambda_2 \lambda^2 -  \left(1 - 20\xi + \xi^2\right) \lambda_3 \lambda^2 - 74\, \xi^2 \lambda^4 \Big].
}

As a rudimentary check, we notice that taking $\xi = 1/N$ makes the contributions to $\delta \lambda_1$ from diagrams (A2) and (A3) exactly cancel, leaving diagram (A1) as the only process contributing to the $\beta$-function for $\lambda_1$; the same non-trivial cancellation occurs in the fundamental CS-matter model. Moreover, in this limit, the counterterms are all the same, and are  given by
\bel{
  \delta \lambda_i = -\frac3{32\pi^2 \eps} \lambda_i \lambda^2.
}
This is precisely the leading $N$ behavior found in \cite{Aharony:2011jz}. This is expected, as each of the three traces in the action \eqref{def S} should collapse to the usual six-point term for fundamental matter, $(\phi^{\dagger} \phi)^3$, and so all three couplings $\lambda_i$ should flow in the same way.

\begin{figure}
  \centering
  \includegraphics[width=\textwidth, keepaspectratio = true]{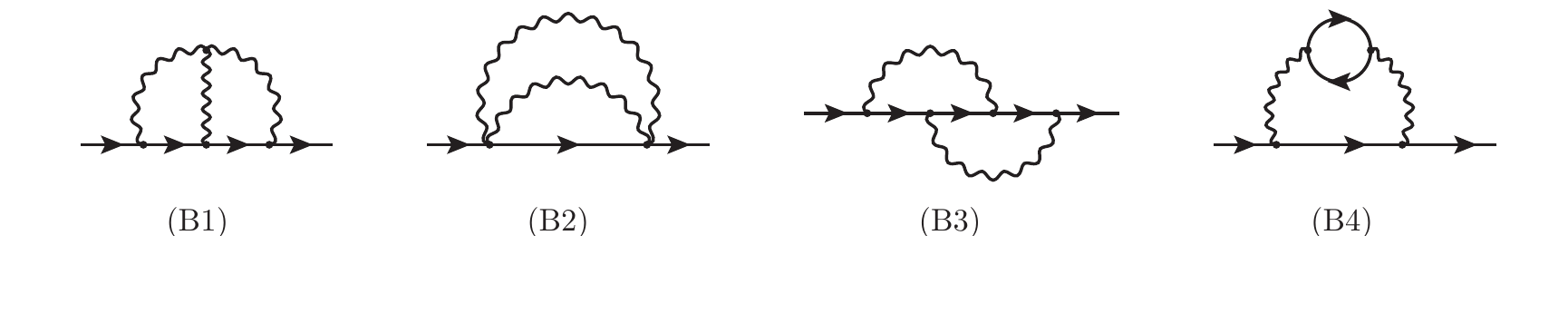}\\
  \caption{\small A schematic depiction of diagrams contributing to the field strength renormalization to two loops.}\label{fig:feyn2loopZ}
\end{figure}

The field strength renormalization can be found by computing two-loop corrections to $\avg{\phi_{ii'}\+ \phi_{jj}}$. These are given by four diagrams on Fig.~\ref{fig:feyn2loopZ}, and these are regulated by the following counterterms \cite{Chen:1992ee}:
\algnl{
  \trm{(B1):}\quad & \delta Z = - \frac{g^4}{24\pi^2} \left(N^2 + M^2 \right) \frac1\eps ,\\
  \trm{(B2):}\quad & \delta Z =  \frac{g^4}{24\pi^2} \left(\frac{N^2 + M^2}4 + 4NM \right)\frac1\eps,\\
  \trm{(B3):}\quad & \delta Z = \frac{g^4NM}{6\pi^2}\frac1\eps,\\
  \trm{(B4):}\quad & \delta Z = \frac{g^4NM}{12\pi^2}\frac1\eps.
}
The total counterterm needed, expressed in terms of 't Hooft couplings, is
\bel{\label{delta Z}
  \delta Z = -\frac{\lambda^2}{96\pi^2\eps} \left(3 - 40\xi + 3\xi^2\right).
}
This result precisely reduces to the leading $N$ result found in \cite{Aharony:2011jz} when $\xi = 1/N$.

The $\beta$-functions, $\beta_i = \mu \del_\mu \lambda_i$, are now found using standard methods. In dimensional reduction (regularization), the scale independence of the bare six-point couplings requires that
\bel{
  \mu \pder{}\mu \left[\mu^{2\eps} \frac{\lambda_i + \delta \lambda_i}{(1 + \delta Z)^3}\right] = 0.
}
This is the renormalization group equation. Differentiating through and setting $\mu \del_\mu \lambda = - \eps \lambda$ (the CS coupling runs with the scale in $d - \eps$ dimensions), we find
\bel{
  \beta_i = -2\eps \left[\delta \lambda_i - \lambda_j \pder{(\delta\lambda_i)}{\lambda_j}  + 3\lambda_i \lambda_j \pder{(\delta Z)}{\lambda_j}\right] + \eps \lambda \left[\pder{(\delta\lambda_i)}\lambda - 3 \lambda_i \pder{(\delta Z)}{\lambda}  \right].
}
Knowing that the counterterms are all quadratic or quartic functions of the couplings simplifies these expressions down to
\bel{
  \beta_i = 2\eps \left(\delta \lambda_i - 3\lambda_i \delta Z \right).
}
Substituting the counterterms \eqref{delta lambda} and \eqref{delta Z} finally yields
%\algnl{
 % \beta_1
 % &= \frac{\xi}{16\pi^2} \left[3\, \lambda_1^2 - 8 \lambda_1 \lambda^2 \right],\\
 % \beta_2
  %&= \frac\xi{64\pi^2} \left[16\, \lambda_1 \lambda_2 + 25 (1 + \xi) \lambda_1 \lambda^2 - 26\, \lambda_2 \lambda^2 \right],\\
  %\beta_3
 % &= \frac{\xi}{32\pi^2} \left[12 \left(\xi \lambda_1^2 + \lambda_2^2 + (1 + \xi) \lambda_1 \lambda_2\right) + 6\xi \lambda_1\lambda^2 + 3 (1 + \xi) \lambda_2 \lambda^2 - 13 \lambda_3 \lambda^2 \right].
%}
\algnl{
  \beta_1
  &= \frac{\xi}{16\pi^2} \left[3\, \lambda_1^2 + 8\lambda_1 \lambda^2 - 72 \lambda^4 \right],\\
  \beta_2
  &= \frac\xi{16\pi^2} \left[3\, \lambda_1^2 (1 + \xi) + 4\, \lambda_1 \lambda_2 + 50 (1 + \xi) \lambda_1 \lambda^2 + 32\, \lambda_2 \lambda^2 + 33(1 + \xi) \lambda^4\right],\\
  \beta_3
  &= \frac{\xi}{8\pi^2} \left[3 \left(\xi \lambda_1^2 + \lambda_2^2 + (1 + \xi) \lambda_1 \lambda_2\right) + 12\, \xi \lambda_1\lambda^2 + 6 (1 + \xi) \lambda_2 \lambda^2 + 10\, \lambda_3 \lambda^2 - 111\, \xi \lambda^4 \right].
}

\begin{figure}
  \centering
  \includegraphics[width = 0.46\textwidth, keepaspectratio = true]{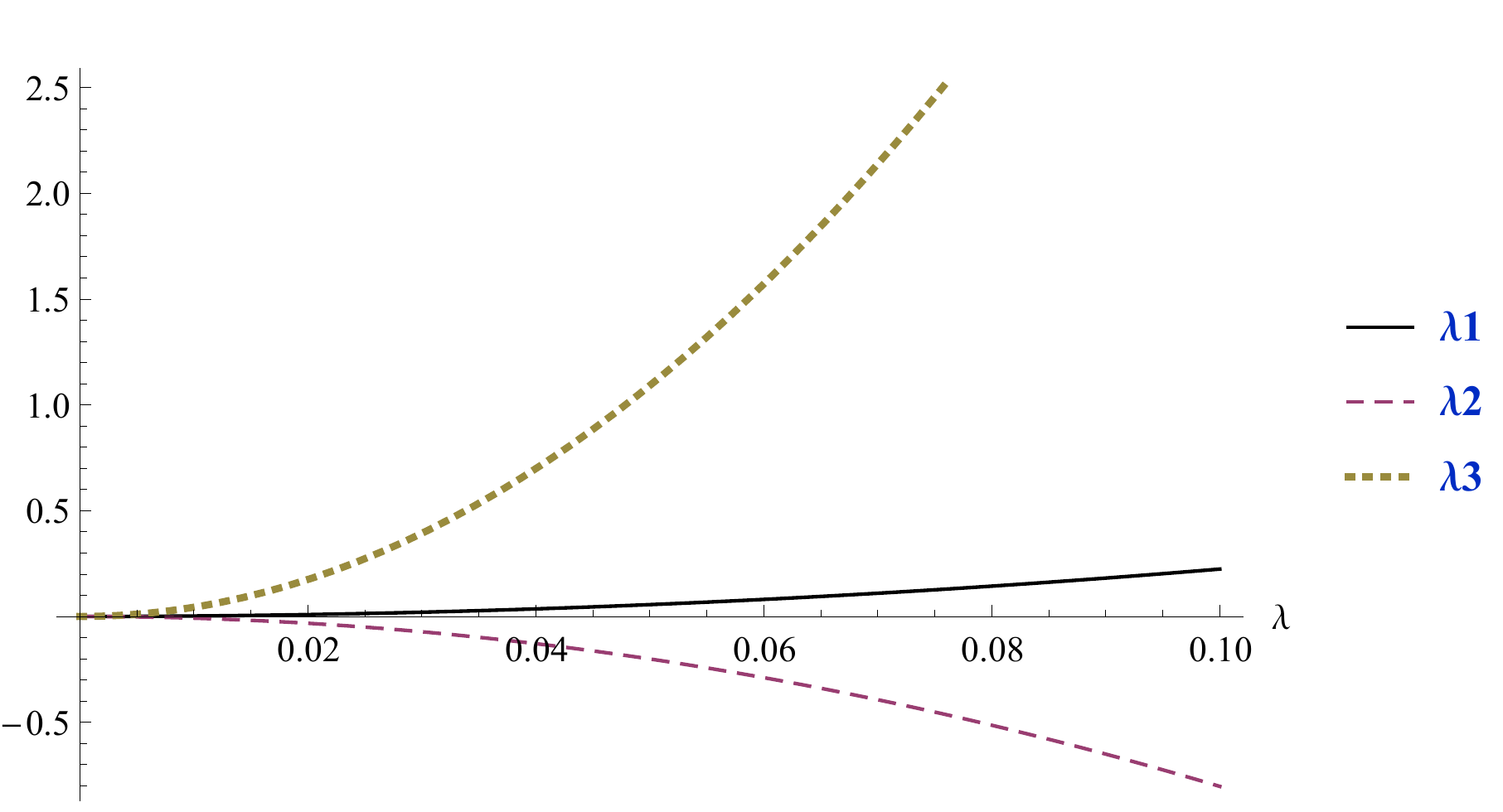} \hspace{2ex}  \includegraphics[width = 0.46\textwidth, keepaspectratio = true]{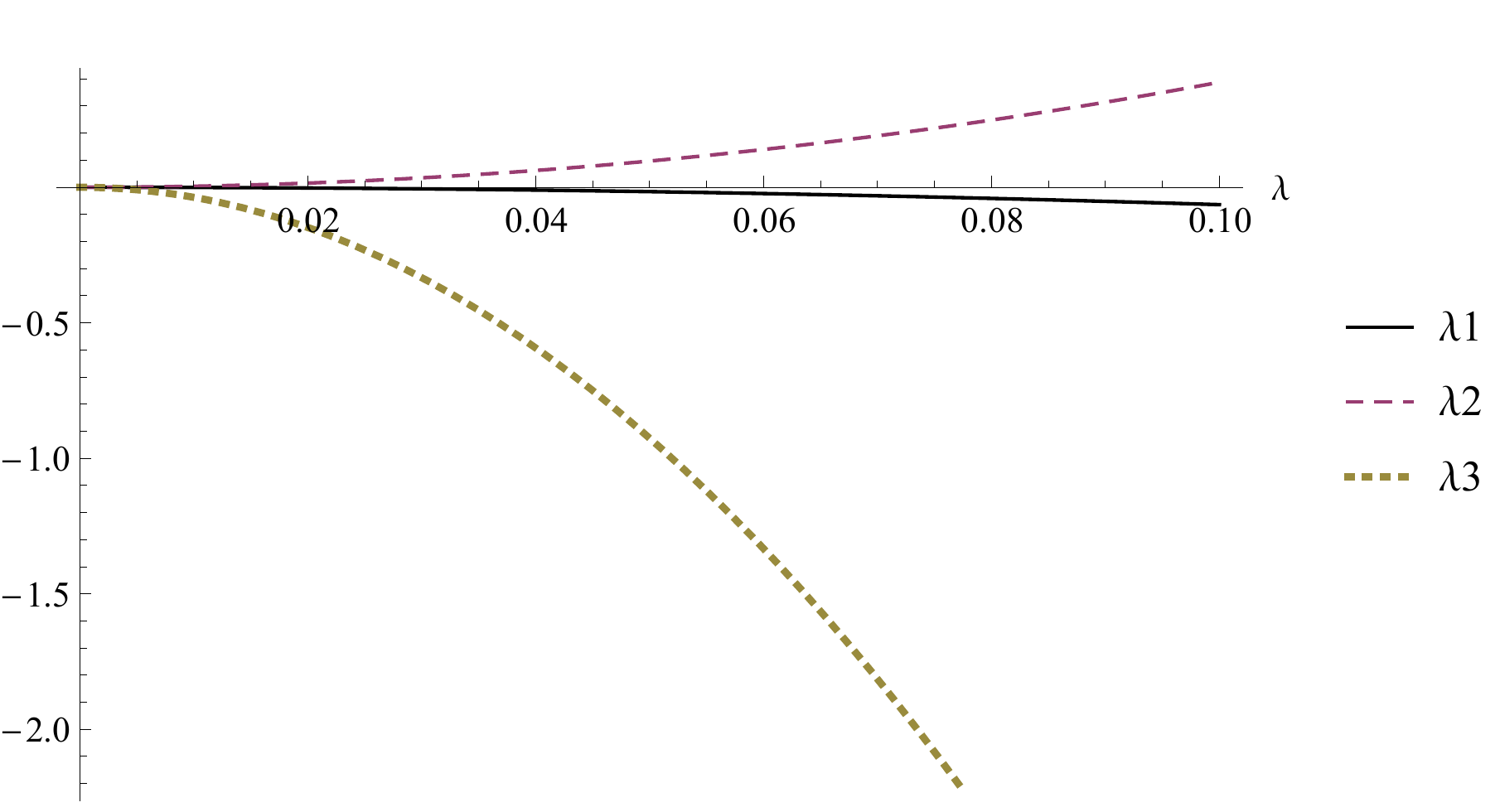}\\
  \caption{\small Two sets of fixed points parametrized by the 't Hooft coupling $\lambda$ in the weak coupling regime at $\xi = 1/2$. The diagram on the left corresponds to choosing the plus sign in eqs.~\eqref{lambda1}--\eqref{lambda3}.}\label{fig:fixedPoints}
\end{figure}

In agreement with \cite{Aharony:2011jz}, we find that the $\beta$-functions all become suppressed when $\xi = 1/N$.  At finite $\xi$, we find two lines of fixed points parametrized by $\lambda$ and given by
%\algnl{
  %\lambda_1^*
  %&= \frac83 \lambda^2,\\
  %\lambda_2^*
 % &= -4(1 + \xi) \lambda^2,\\ % = -25\frac{(1 + \xi)(8 + \xi)}{50 + 4\xi} \lambda^2,\\
 % \lambda_3^*
 % &= - \frac{64}{13}\left(33 - 14\xi + \xi^2\right) \lambda^2.
%}
\algnl{\label{lambda1}
  \lambda_1^*
  &= \frac23 \left(-2 \pm \sqrt{58}\right) \lambda^2,\\ \label{lambda2}
  \lambda_2^*
  &=  -\frac{21}8 \frac{4 \sqrt{58} \pm 7}{\sqrt{58} \pm 10 } \lambda^2 (1 + \xi), \\ \label{lambda3}
  \lambda_3^*
  &= - \frac1{1920} \frac{ 63\left(13103 \mp 2056 \sqrt{58}\right) (1 + \xi^2) + \left(900002 \mp 348400 \sqrt{58}\right) \xi}{(\sqrt{58} \pm 10 )^2} \lambda^2.
}
These two lines of fixed points are shown on Fig.~\ref{fig:fixedPoints}.

We have found this two-loop fixed line in the large-$N$ limit, but our analysis holds for any value of the fixed ratio $\xi = M/N$. It will be interesting to see if this holds to all-loop order, at least in the large-$N$ limit. We can see that as the 't Hooft coupling goes to zero the fixed points approach the trivial value zero, which is consistent with the fact that that the model has no fixed points in the absence of the CS term.

%If we take the fixed point value of the single trace operator to be positive then the fixed point values of the other multiple trace operators are negative and vice-versa. One may wonder if the system is stable or not. To answer that question one has to compute the Coleman-Weinberg potential and see if the potential is bounded from below for the specific values of the coupling constants. We will not try to do that in this paper, but an argument may be given which shows that there is no tachyonic mode at least for small values of the 't Hooft coupling $\lambda$. For example, one could compute the self-energy of the scalar field and look for unphysical poles. In our case the self energy at least up to two loop order does not get any contribution form the six-point couplings and so there cannot be a tachyonic pole caused by the negative coupling constants of the six-point interactions.

It is easy to see that, at both fixed point lines, the sign of the $\lambda_2$ coupling is opposite that of the other two couplings. Both lines could be stable. To answer that question one has to compute the Coleman-Weinberg potential and see if the potential is bounded from below for the specific values of the coupling constants. We will not try to do that in this paper, but an argument may be given which shows that there is no tachyonic mode --- at least for small values of the 't Hooft coupling $\lambda$. For example, one could compute the self-energy of the scalar field and look for unphysical poles. In our case the self-energy at least up to two-loop order does not get any contribution from the six-point couplings, and so there cannot be a tachyonic pole caused by the negative coupling constants of the six-point interactions.

In addition, the conformal symmetry may be spontaneously broken. For example, the conformal symmetry is spontaneously broken in the vector model with only a $\phi^6$ interaction and no CS term, if the coupling constant $\lambda_6$ is greater than $4\pi^2$. It will be interesting to compute the Coleman-Weinberg potential and study this spontaneous breaking, but we leave that for future research.

\section{No light states on a torus} \label{sec:torus}

In this section we change gears and study the low-energy, small-$\lambda$ limit of the CS-bifundamental theory on the torus in a spacetime with Lorentzian signature. Our goal is to retrieve the spectrum of the theory; our methods closely follow those in \cite{Banerjee:2012gh}. We will see that this spectrum has no state whose energy vanishes in the 't Hooft limit. In principle, one should study the full bifundamental theory with its potential terms. Instead, we will study the toy theory with only the CS terms and the covariant kinetic terms for the scalar field. This captures the essential physics at small 't Hooft coupling and at energy low compared to the inverse size (KK scale) of the spatial torus.

\subsection{Low-energy effective Hamiltonian}

As usual, we canonically quantize in the gauge $A_0 = B_0 = 0$, where $0$ denotes the time component. We are interested in low-energy degrees of freedom only, and so we disregard the spatially varying (non-zero momentum) modes of all of the fields, as these necessarily have a gap set by the size of the torus.\footnote{See \cite{Banerjee:2012gh} for a detailed discussion on the justification of this process.} This is just dimensional reduction onto a single spatial point; naturally, the dimensionally reduced theory is just a form of quantum mechanics. Its Lagrangian can be written as
\bel{
  L = \frac k{2\pi} \Tr \left(A_1 \dot A_2 - B_1 \dot B_2\right) + \Tr \left(\dot \phi\+ \dot \phi - \phi\+ A^2 \phi - \phi B^2 \phi\+ + 2 A_i \phi B_i \phi\+ \right).
}
Note that the minus signs in the $\phi^2 A^2$ terms come from our choice of the metric convention (mostly minus). From here on, we treat all variables ($A_i$, $B_i$, and $\phi$, with $i = 1, 2$ and $A^2 = A_i A_i = A_1^2 + A_2^2$) as matrices of c- or q-numbers, and we drop the indices on the traces. Moreover, to compactify notation, we write the sum of traces as a trace of a sum, even when the matrices in this sum are not all of the same dimension.

In \cite{Banerjee:2012gh}, a simpler version of this model --- the one arising from the theory of fundamental scalars coupled to one gauge field --- was studied by canonical quantization, treating the $\phi\+ A_i^2 \phi$ term perturbatively in $1/N$. This method readily gives the spectrum of the theory, and we will follow the same approach here. We will also develop a formal justification for using perturbation theory.

Letting $g^2 \equiv 4\pi/k$, we choose the canonical variables defined by
\bel{
  A_1 \equiv \frac g{\sqrt 2} P_a T^a,\ A_2 \equiv \frac g{\sqrt 2}  Q_a T^a,\quad B_1 \equiv \frac g{\sqrt 2}  P_\alpha T^\alpha,\ B_2 \equiv - \frac g{\sqrt 2}  Q_\alpha T^\alpha.
}
We use $a$, $b$, etc.~to denote generators of $SU(N)$, and $\alpha$, $\beta$, etc.~to denote generators of $SU(M)$. The Hamiltonian is
\algnl{\notag
  H
  &= \Tr\left(\pi\+ \pi\right) + \frac12 g^2 \left(P_a P_b + Q_a Q_b\right) \b M^{ab}(\phi)\ +\\
  &\qquad + \frac12 g^2 \left(P_\alpha P_\beta + Q_\alpha Q_\beta\right) \b M^{\alpha\beta}(\phi) + g^2 \left(P_a P_\alpha - Q_a Q_\alpha\right) \b M^{a \alpha}(\phi),
}
with
\algnl{
  \b M^{ab}(\phi)
  &\equiv \Tr \left(\phi\+ T^a T^b \phi \right),\\
  \b M^{\alpha \beta}(\phi)
  &\equiv \Tr \left(\phi T^\alpha T^\beta \phi\+ \right),\\
  \b M^{a\alpha}(\phi)
  &\equiv \Tr \left(T^a \phi T^\alpha \phi\+ \right).
}

It is useful to switch to variables that will give the ladder operators upon quantization, and so we let
\algnl{
  P_a \equiv \frac1{\sqrt 2} \left(c_a\+ + c_a\right),&\quad Q_a \equiv \frac1{i\sqrt 2}\left(c_a\+ - c_a\right),\\
  P_\alpha \equiv \frac1{\sqrt 2} \left(d_\alpha\+ + d_\alpha\right),&\quad Q_\alpha \equiv \frac1{i\sqrt 2}\left(d_\alpha\+ - d_\alpha\right).
}
The Hamiltonian becomes
\algnl{\notag
  H
  &= \Tr\left(\pi\+ \pi\right) + \frac12 g^2 \left(c_a c_b\+ + c_a\+ c_b \right) \b M^{ab}(\phi)\ +\\
  &\qquad + \frac12 g^2 \left(d_\alpha d_\beta\+ + d_\alpha\+ d_\beta\right) \b M^{\alpha\beta}(\phi) + g^2 \left(c\+_a d\+_\alpha + c_a d_\alpha\right) \b M^{a \alpha}(\phi).
}
Notice that the choice of a negative level for the $SU(M)$ CS action translates into the last term of the Hamiltonian above. This term allows for simultaneous creation of $SU(N)$ and $SU(M)$ holonomies. In other words, the conserved quantum number of ``particles'' created by the $c_a$'s and $d_\alpha$'s will be the difference (rather than the sum) of numbers of ``particles'' of each species.

To quantize the holonomy degrees of freedom, we impose
\bel{
  [c_a, c_b\+] = \delta_{ab},\quad [d_\alpha, d_\beta\+] = \delta_{\alpha\beta}.
}
Using $T^aT^a = C_2(N)$ for $SU(N)$ (or for any simple group) and likewise for $SU(M)$, the normal-ordered Hamiltonian becomes
\algnl{\notag
  H
  &= \Tr\left(\pi\+ \pi\right) + \frac12 g^2 (C_2(N) + C_2(M)) \;\Tr\left(\phi\+\phi\right) + \\
  &\qquad + g^2\; c_a\+ c_b \; \b M^{ab}(\phi)\ + g^2\; d_\alpha\+ d_\beta\; \b M^{\alpha\beta}(\phi) + g^2 \left(c\+_a d\+_\alpha + c_a d_\alpha\right) \b M^{a \alpha}(\phi),
}
and we see that the scalar fields will acquire a mass  due to the vacuum energy of the holonomies. This mass is set by
\bel{
  m^2 \equiv \frac12 g^2 (C_2(N) + C_2(M)) = \frac{2\pi}k (C_2(N) + C_2(M)).
}

The mass of the holonomy excitations will be set by the vacuum energy of the scalar excitations, and to find it we must quantize the scalars as well. We let
\bel{
  \pi \equiv \sqrt\frac m2 \left(b\+ + a \right),\quad \phi \equiv \frac 1{i\sqrt{2 m}} \left(b\+ - a\right),
}
where $a$ and $b$ are, respectively, $N\times M$ and $M \times N$ q-number matrices. The conjugation operation ${}\+$ acts on such a matrix by transposing it and taking a Hermitian conjugate of each q-number element. We impose cannonical commutation relations on each of the $NM$ elements of $a$ or $b$, and so instead of the standard cyclicity of the trace we have
\bel{
  \Tr\left(a a\+ \right) = \Tr\left(a\+ a\right) + NM.
}
Similarly, for an $M\times M$ matrix $T$, we find
\bel{
  \Tr\left(a T a\+ \right) = \Tr\left(T a\+ a\right) + N\, \Tr\, T
}
and, for an $N \times N$ matrix $S$,
\bel{
  \Tr\left(S a T a\+ \right) = \Tr\left(T a\+ S a \right) + (\Tr\, T)(\Tr\, S).
}
These identities allow us to find the normal-ordered form of the Hamiltonian, which is
\algnl{\notag
  H
  &=  m\, \Tr\left(b\+ b + a\+ a \right) + m NM  \\
  &\qquad + g^2\; c_a\+ c_b \; \b M^{ab}(a, b)\ + g^2\; d_\alpha\+ d_\beta\; \b M^{\alpha\beta}(a, b) + g^2 \left(c\+_a d\+_\alpha + c_a d_\alpha\right) \b M^{a \alpha}(a, b),
}
where, using $\Tr\left(T^a T^b\right) = C(N)\delta^{ab}$,
\algnl{
  \b M^{ab}(a, b)
  &= \frac{M}{2m}C(N)\delta^{ab} + \frac1{2m} \Tr \left(a\+ T^a T^b a + T^a T^b b\+  b - b T^a T^b a - a\+ T^a T^b b\+ \right),\\
  \b M^{\alpha \beta}(a, b)
  &= \frac{N}{2m}C(M)\delta^{\alpha\beta} + \frac1{2m} \Tr \left(T^\alpha T^\beta a\+ a + b\+ T^\alpha T^\beta b - a T^\alpha T^\beta b - b\+ T^\alpha T^\beta a\+ \right),\\
  \b M^{a\alpha}(a, b) \label{M-aalpha}
  &= \frac{1}{2m}(\Tr\, T^a)(\Tr\, T^\alpha) + \frac1{2m} \Tr \left(T^\alpha a\+ T^a a + T^a b\+ T^\alpha b  - T^a b\+ T^\alpha a\+ - T^a a T^\alpha b \right).
}

\subsection{The unperturbed spectrum}

The spectrum is now easily found using perturbation theory. We shift the ground state energy to zero and choose the unperturbed Hamiltonian to be
\algnl{\notag
  H_0
  &=  m\, \Tr\left(b\+ b + a\+ a \right) + m_{SU(N)} \Tr\left(c\+ c\right) + m_{SU(M)} \Tr\left(d\+ d\right) +\\ \label{def H0}
  &\qquad + \frac{g^2}{2m} \Tr\left(a\+ c\+ c a + c\+ c b\+ b + d\+ d a\+ a + b\+ d\+ d b \right).
}
This Hamiltonian can be exactly diagonalized in the 't Hooft limit. The Hilbert space is spanned by the standard SHO eigenstates of the quadratic part of $H_0$, and we will later show that the quartic term merely provides a correction to some of the eigenenergies. Hence, we will from now on adopt the usual language of creation/annihilation operators. The term that we have to treat perturbatively, $V \equiv H - H_0$, is
\bel{\label{def V}
  V
  = -\frac{g^2}{2m}\Tr\left(b c\+ c a + a\+ c\+ c b\+ + ad\+ db + b\+ d\+ d a \right) + \frac{g^2}{2m} \left(c\+_a d\+_\alpha + c_a d_\alpha \right) \b M^{a\alpha}(a, b),
}
where $\b M^{a \alpha}(a, b)$ is defined in eq.~\eqref{M-aalpha}. We have already disregarded the terms proportional to the traces of group generators, as these are zero for the case at hand.

The unperturbed Hamiltonian contains a mass term for the holonomy degrees of freedom. For any gauge group, one set of holonomy excitations has a bare mass of $g^2 M C(N)/2m$.\footnote{As already hinted, this mass will be corrected by some of the quartic terms, and hence we refer to the coefficients of the quadratic terms in $H_0$ as bare masses.} In our case, the generators $T^a$ are in the fundamental representation of $SU(N)$, for which $C(N) = 1/2$ and $C_2(N) = (N^2 - 1)/2N$. In the 't Hooft limit ($N \rar \infty$ while $N/k = \lambda$ and $M/N = \xi$ are held fixed) the bare mass is
\bel{
  m_{SU(N)} = \frac \xi{1 + \xi} m  \propto \sqrt\lambda\frac \xi{\sqrt{1 + \xi}}.
}
We approach the regime of CS coupled to fundamental matter by taking $\xi \rar 0$, and in this limit the holonomy states become light, as found in \cite{Banerjee:2012gh}. On the other hand, the states of the other holonomy have bare mass $g^2 NC(M)/2m$, and in the 't Hooft limit this is
\bel{
  m_{SU(M)} = \frac 1{1 + \xi} m  \propto \sqrt\lambda \frac1{\sqrt{1 + \xi}}.
}
These states remain massive as we approach the fundamental representation by letting $\xi$ become $O(1/N)$.

The physical eigenstates of the unperturbed Hamiltonian must be invariant under the remaining $SU(N) \times SU(M)$ gauge transformations. Thus, if $\qvec \Omega$ is the vacuum of the theory, $c_a\+ \qvec\Omega$ is not a physical state, but $\Tr\left(c\+\right)^2$ is, and so is any state created by a gauge-invariant combination of creation operators.\footnote{The state $\Tr\left(c\+\right)\qvec\Omega$ is not physical because the generators of $SU(N)$ are traceless.} The operators we consider transform as
\bel{
  c\+ \mapsto U c\+ U\+,\quad d\+ \mapsto V d\+ V\+,\quad a\+ \mapsto V a\+ U\+,\quad b\+ \mapsto U b\+ V\+.
}
(Recall that $c\+$ and $d\+$ are $N\times N$ and $M \times M$ matrices, while $a\+$ is $M \times N$ and $b\+$ is $N \times M$.) The following single-trace operators are all gauge-invariant:
\bel{
  \mathcal C_n \equiv \Tr\left(c\+\right)^n,\quad \mathcal D_n \equiv \Tr\left(d\+\right)^n,\quad \mathcal E_n^{\vec k,\, \vec \ell} \equiv \Tr\left[\prod_{i = 1}^n a\+ \left(c\+\right)^{k_i} b\+ \left(d\+\right)^{\ell_i} \right].
}
The vectors $\vec k$ and $\vec \ell$ have $n$ components each, and it is understood that permuting the entries of either one does not generate a physically new state. Furthermore, due to the relations between traces of matrix powers, not all single-trace operators are independent. For instance, out of the holonomy degrees of freedom, only $\mathcal C_2, \ldots, \mathcal C_N$ and $\mathcal D_2, \ldots, \mathcal D_M$ are independent; the others can all be written as multi-trace combinations of these operators. The interdependence of these operators is a finite-$N$ or high-energy effect, and it will not figure in our analysis of the low-energy spectrum.

The states created by these operators must all have unit norm. In the 't Hooft limit, one can check that the correct normalization is
\algnl{\label{c-norm}
  \qvec{\mathcal C_n}
  &= \frac{2^{n/2}}{\sqrt{n}} \frac1{N^{n/2}} \mathcal C_n \qvec{\Omega},\\ \label{d-norm}
  \qvec{\mathcal D_n}
  &= \frac{2^{n/2}}{\sqrt{n}} \frac1{M^{n/2}} \mathcal D_n \qvec{\Omega},\\ \label{e-norm}
  \qvec{\mathcal E_n^{\vec k,\, \vec \ell}}
  &= \frac{2^{(k + \ell)/2}}{\sqrt{s}} \frac1{N^{(k + n)/2} M^{(\ell + n)/2}} \mathcal E_n^{\vec k,\, \vec \ell} \qvec{\Omega}, \quad k \equiv \sum_{i = 1}^n k_i,\ \ell \equiv \sum_{i = 1}^n \ell_i.
}
Above we use $s$ to denote the ``symmetry factor,'' with $s = n$ if $\ell_i = \ell/n$ and $k_i = k/n$ for all $i$, and with $s = 1$ otherwise. These normalizations can be derived using planar diagram techniques that we will introduce below, and then we will also explain why one may think of each gauge boson as contributing a factor of $1/\sqrt N$ or $1/\sqrt M$ to the normalization of a state, while each  pair of scalar excitations contributes $1/\sqrt{NM}$.

We will now demonstrate that the states \eqref{c-norm}, \eqref{d-norm}, and \eqref{e-norm} are approximate eigenstates of the Hamiltonian $H_0$, given by \eqref{def H0}. Focus on the action of the quartic terms on these states. Each of the quartic terms annihilates any $\qvec{\mathcal C_n}$ or $\qvec{\mathcal D_n}$, so states formed from purely gauge excitations have unperturbed masses equal to their bare masses. (The lightest state in the spectrum, $\qvec{\mathcal C_2}$, falls in this group, and has bare/unperturbed energy $2m_{SU(N)}$.) Similarly, states $\qvec{\mathcal E_n^{0,0}}$ with only matter excitations preserve their bare mass once the quartics in $H_0$ are introduced. All other states will be non-trivial eigenstates of the quartic operators in $H_0$, at least at large $N$. For instance, $\qvec{\mathcal E_1^{1, 0}}$ is an exact eigenstate of $H_0$. A short computation shows that
\algnl{\notag
  \Tr\left(a\+ c\+ c a\right) \qvec{\mathcal E_1^{1, 0}}
  &= \mathcal N(\mathcal E_1^{1, 0})\ \Tr\left(a\+ c\+ c a\right)\Tr\left(a\+ c\+ b\+\right) \qvec\Omega \\ \notag
  &= \mathcal N(\mathcal E_1^{1, 0})\ \Tr\left(a\+ c\+ c  c\+ b\+\right) \qvec\Omega \\
  &= \frac N2 \qvec{\mathcal E_1^{1, 0}}.
}
Similarly, $\qvec{\mathcal E_1^{1, 0}}$ is an eigenstate of $\Tr\left(c\+ c b\+ b \right)$ with the same eigenvalue, and so
\bel{
  H_0 \qvec{\mathcal E_1^{1, 0}} = \left(2m + m_{SU(N)} + \frac{g^2}{2m} N \right) \qvec{\mathcal E_1^{1, 0}} =  \frac{4 + 3\xi}{1 + \xi} m \qvec{\mathcal E_1^{1, 0}}.
}

In general, it is true that
\bel{
  \frac{g^2}{2m}\Tr\left(a\+ c\+ c a\right)\qvec{\mathcal E_n^{\vec \ell,\, \vec k}} =  \frac{g^2}{2m} \Tr\left(c\+ c b\+ b \right)\qvec{\mathcal E_n^{\vec \ell,\, \vec k}} = n(1 - \delta_{\ell, 0}) \frac{m}{1 + \xi}  \qvec{\mathcal E_n^{\vec \ell,\, \vec k}} + O\left(\frac1N\right)
}
and
\bel{
  \frac{g^2}{2m} \Tr\left(d\+ d a\+ a\right)\qvec{\mathcal E_n^{\vec \ell,\, \vec k}} = \frac{g^2}{2m} \Tr\left(b\+ d\+ d b \right)\qvec{\mathcal E_n^{\vec \ell,\, \vec k}} = n(1 - \delta_{k, 0}) \frac{\xi m}{1 + \xi} \qvec{\mathcal E_n^{\vec \ell,\, \vec k}} + O\left(\frac1N\right).
}
The $1/N$ corrections come from the multi-trace operators generated by the action of the quartics on some of the $n > 1$ states. For instance, take $n = 2$ and $\vec \ell = (1, 1)$; we find
\bel{
  \Tr\left(a\+ c\+ c a\right) \qvec{\mathcal E_2^{(1, 1),\, 0}} = N\qvec{\mathcal E_2^{(1, 1),\, 0}} + \frac{\mathcal N(\mathcal E_2^{(1, 1),\, 0})}{\mathcal N^2(\mathcal E_1^{1,0})}\qvec{\mathcal E_1^{1,0}, \mathcal E_1^{1, 0}},
}
where we use $\mathcal N(\mathcal O)$ to denote the normalization of the state created by $\mathcal O$, as shown in eqs.~\eqref{c-norm}, \eqref{d-norm}, and \eqref{e-norm}. The second term has an $O(1)$ coefficient that is negligible in the 't Hooft limit.

The upshot of these calculations is that the unperturbed Hamiltonian at large $N$ and $M$ has the same eigenstates as the quadratic, SHO Hamiltonian, with unperturbed energies
\algnl{\notag
  E_0(\mathcal C_n)
  &= n m_{SU(N)},\\
  E_0(\mathcal D_n)
  &= n m_{SU(M)},\\ \notag
  E_0(\mathcal E_n^{\vec \ell,\, \vec k})
  &= 2nm + \left[ \ell + 2n(1 - \delta_{\ell, 0})\right] m_{SU(N)} + \left[ k + 2n(1 - \delta_{k, 0})\right] m_{SU(M)}.
}

\subsection{Perturbative corrections}

The remaining quartic couplings, assembled in \eqref{def V}, can be treated perturbatively. We wish to find corrections $E_n$ to energies of the non-degenerate eigenstates of $H_0$. Calculating $E_n$'s at $n > 2$ can in principle be done by the usual methods of quantum-mechanical perturbation theory.  However, studying all possible contractions between traces quickly becomes a very involved task.

\begin{figure}
 \centering
 \includegraphics[keepaspectratio=true, scale = 0.8, trim= 45pt 60pt 15pt 80pt]{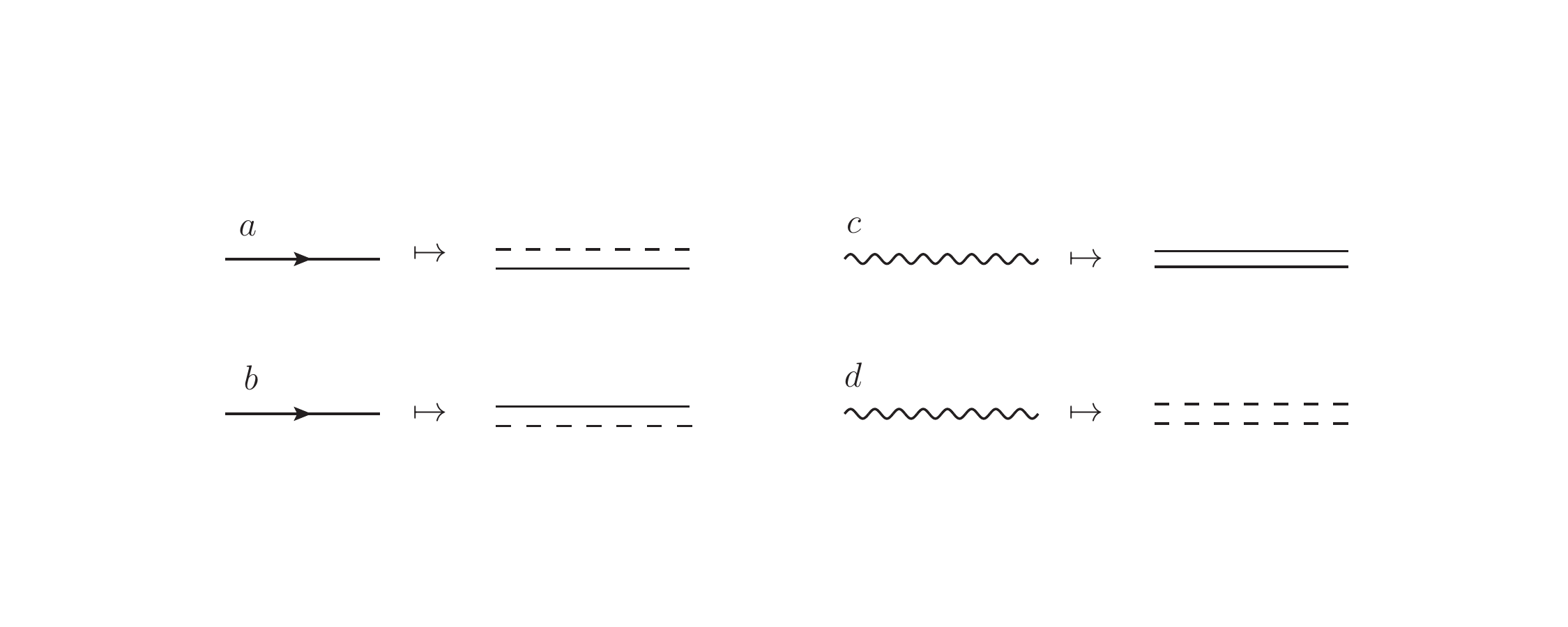}
 % doubleline.pdf: 0x0 pixel, 0dpi, 0.00x0.00 cm, bb=
 \caption{\small Converting to double-line notation. Dashed lines correspond to indices running from $1$ to $N$, and full lines correspond to indices running from $1$ to $M$.}
 \label{fig:doublelines}
\end{figure}

We will now show that, fortunately, perturbative calculations drastically simplify by using a diagrammatic technique that automatically keeps track of index contractions in expressions like $\qmat{s_1}V{s_2}\qmat{s_2}V{s_3}\qmat{s_3}V{s_1}$. We do not rederive quantum mechanical perturbation theory; we merely note that known expressions for perturbative corrections can be efficiently encoded using diagrams. The ``Feynman rules'' for computing energy corrections are as follows:

\begin{enumerate}
 \item The perturbatively corrected energy $E(\O)$ of a state $\qvec \O$ can be represented as a Feynman diagram computation of the self-energy of this state. Each of the particle species ($a$, $b$, $c$ and $d$) is assigned a distinct line in these diagrams.
 \item Draw all diagrams that have the particles comprising $\qvec \O$ in both the in- and the out-state. Each interaction is represented by a four-point vertex, see Fig.~\ref{fig:vertices}. The amplitude $\qmat{\trm{out}}V{\trm{in}}$ contains only one vertex. The $n^{\trm{th}}$ order correction to the energy, $E_n(\O)$, comes from diagrams with  $n$ vertices, with the sum over intermediate states replaced by the sum over \emph{planar}, \emph{connected} diagrams. The planar diagrams are those in which no lines cross; when drawing diagrams, it is important to note that diagrams can be planar up to a cyclic rearrangement of lines coming out of a vertex.

\begin{figure}[tb]
 \centering
 \includegraphics[scale = 0.8, trim=45pt 80pt 15pt 100	pt]{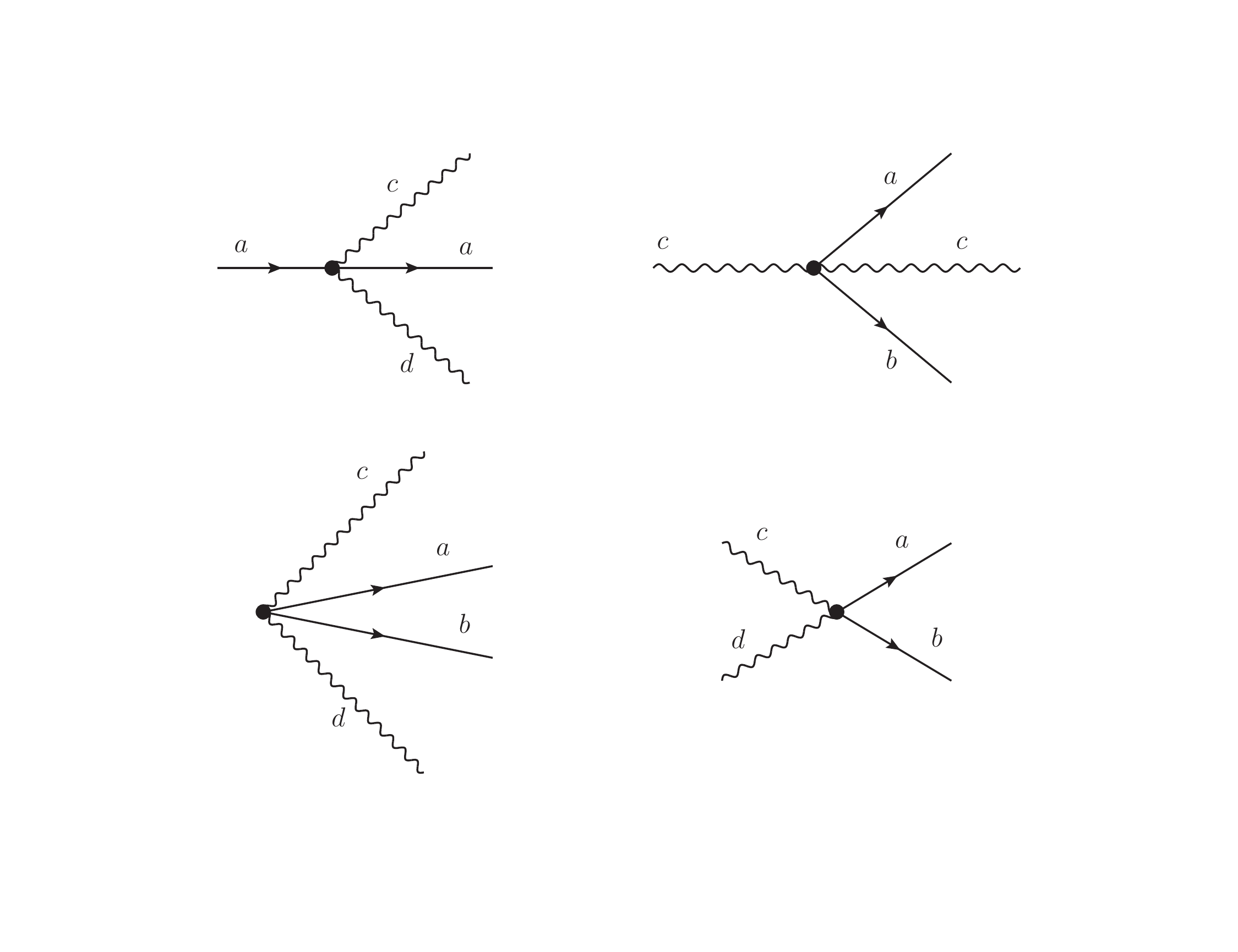}
 \caption{\small Some of the interactions that appear in perturbation theory, as follows from the perturbation potential $V$ in eq.~\eqref{def V}. Time flows from left to right, as indicated by the arrows on the scalars. All other allowed vertices are found by switching the time direction or by exchanging $c \leftrightarrow d$ or $a \leftrightarrow b$. For the purposes fo counting planar diagrams, it is crucial to preserve the correct ordering of lines that emerge from each vertex (not shown in the figure); these must match the order of appearance in the trace contained in $V$, and one is only allowed to cyclically permute them.}
 \label{fig:vertices}
\end{figure}

 \item For each vertex in a given diagram, write a factor of $g^2/2m$ with the sign that appears in eq.~\eqref{def V}.
 \item For each \emph{internal} gauge boson contraction in a given diagram, write a factor of $1/2$. This is because the gauge boson propagator is found through
 \bel{
   \avg{c_{ij} (c\+)_{kl}} = \avg{c_a c_b\+} T^a_{ij} T^b_{kl} = T^a_{ij} T^a_{kl} + \trm{(norm.~ord.)} = \frac12 \delta_{ij}\delta_{kl} + \trm{(norm.~ord.)} + O\left(\frac1N\right).
  }
 The scalar propagator has no corresponding prefactors, so we can just contract scalars with each other without further worries. We must only contract particles of one type with other particles of the same type.
 \item For each amplitude $\qmat{\trm{out}}V{\trm{in}}$, write a factor of $1/\sqrt{s\_{in} s\_{out}}$ to account for the correct normalization of each intermediate state. Here $s$ is the symmetry factor introduced in eq.~\eqref{e-norm}.
 \item For each intermediate state $\qvec \S$ different from the incoming state $\qvec \O$, write a factor of $1/(E_0(\O) - E_0(\S))$. This allows us to mimic the results from perturbation theory.
 \item Each time an intermediate state is $\qvec \O$, i.e.~when we encounter a one-particle-reducible diagram, write a factor of $-1$. If $\qvec \O$ appeared $k$ times as an intermediate state, take the non-zero energy differences $E(\O) - E(\S_i)$, write all possible products of $n - k - 1$ of these differences, sum the inverses of each product, and multiply the total amplitude by this factor.
 \item Finally, and most importantly, write down the factors of $N$ and $M$ that come from internal indices that got contracted. These can be found by converting each Feynman diagram into a double-line diagram \`a la 't Hooft (see Fig.~\ref{fig:doublelines}) and counting the number of loops associated to each of the two types of indices. These loop numbers are the powers of $N$ and $M$ that must be written for each diagram. It is important to note that we use the double-line notation only to count the relevant powers of $N$ and $M$ and \emph{not} to find all possible diagrams and compute them; if we applied the typical double-line analysis of 't Hooft, we would end up computing diagrams that do not represent processes in perturbation theory.
\end{enumerate}

These rules allow for a quick estimate of the size of the effect of each order in perturbation theory. To do so, notice that each energy is of order $m \sim \sqrt\lambda$, while each vertex contributes a factor of order $\sqrt\lambda/N$. At order $n$, the product of these factors gives $\sqrt\lambda/ N^n$. On the other hand, a connected diagram in double-line notation can only have up to $n$ loops in total.\footnote{
  This is so because each vertex contributes a total of eight outgoing lines in double-line notation, and four of these must be ``wasted'' on ensuring that the diagram is actually connected. This leaves four lines per vertex. Looking at the allowed vertices, we find that the new lines emanating from each vertex must be a pair of $N$-lines and a pair of $M$-lines. There are no loops involving just one vertex allowed by the theory, so each vertex must pair up with at least one more vertex to give a loop. Thus, there can be no more than one loop per vertex.
} Any diagrams with fewer than $n$ loops will be suppressed in the 't Hooft limit, so they may be discarded. If a diagram has exactly $n$ loops, some of these loops will be $M$-loops and some will be $N$-loops. In the case of interest to us (i.e.~when computing corrections to $E(\mathcal C_2)$, which we will do in the next section in order to find the gap in the theory), studying the allowed vertices shows that a diagram of order $2n$ will have $n$ loops of each type, while a diagram of order $2n + 1$ will have $n + 1$ $M$-loops and $n$ $N$-loops. Thus, perturbation theory gives an expansion in powers of $\xi \equiv M/N$, with perturbations of orders $2n - 1$ and $2n$ both being suppressed by $\xi^n$. This shows that perturbation theory is well-defined at low energies (where the $1/N$ corrections cannot be compensated for by pure combinatorics, i.e.~where $n \ll N$). In particular, in the case of fundamental matter coupled to CS theory, $\xi$ becomes of order $1/N$, and we recover the perturbation in powers
of $1/N$ found in \cite{Banerjee:2012gh}.

\subsection{State normalizations and vacuum amplitudes}

The rules above are a bit formal and perhaps unintuitive. As already mentioned, the double-line formalism can be used to simply derive the normalizations of physical states, as given in eqs.~\eqref{c-norm}-\eqref{e-norm}. As an illustration of our diagrammatic approach --- and as a means of justifying the specifics of our Feynman rules --- we now derive these normalizations and explain the origin of the symmetry factors $s$.

Consider any gauge-invariant creation operator $\O$. Let the normalized state created by this operator be $\qvec\O = \mathcal N \O \qvec\Omega$. The normalization condition is
\bel{
  \qprod \O\O = \mathcal N^2 \qmat\Omega{\O\+\O}\Omega = 1.
}
How is this condition to be achieved?

The first expectation value, $\qprod\O\O$, is given by the zero-vertex diagram with particles from $\qvec\O$ being both in- and out-going states. Following the Feynman rules we formulated, we find that there are precisely $s$ planar diagrams, where $s$ is the symmetry factor introduced earlier. Only these diagrams need be taken into account. There are no vertices, intermediate states, or internal gauge boson contractions, so each diagram is just given a value of $1/\sqrt{s\cdot s}$ as per rule 5. The symmetry factors cancel each other out, and this shows that the way we have defined the symmetry factor and Feynman rules precisely ensures that each gauge-invariant state is normalized.

The second expectation value is a vacuum amplitude, and the normalization factor is given as the inverse square root of this amplitude. Now $\O$ acts as an interaction term, and all it does is insert a vertex that creates all the particles in $\qvec\O$ with their indices properly contracted in the past. The diagram giving $\qmat\Omega{\O\+\O}\Omega$ can be drawn by using the diagram for $\qprod \O\O$, focusing on the external states, and connecting all adjacent lines in the double-line notation, with the first line being connected to the last one. These connections correspond to index contractions as they appear in the trace. Each boson contraction is now an internal one, and this accounts for the factors of two appearing in the normalization. Each incoming particle now also gives rise to a loop (in the 't Hooft limit), and these account for the factors of $N$ and $M$ in the normalization. (In particular, our heuristic $N$-counting is justified; each pair of $a$ and $b$ matter particles contributes $NM$ to
the vacuum amplitude, each $c$ boson contributes $N$, and each $d$ boson contributes $M$.)  Finally, it is now apparent that there can be only $s$ diagrams that contribute at leading order in $N$; only diagrams that are planar or that can be made planar by cyclic permutations contribute, and there are exactly $s$ of these. In short, we conclude that the vacuum amplitude is
\bel{
  \qmat\Omega{\O\+\O}\Omega = s \cdot (NM)^{\#(ab)} \cdot N^{\#(c)} \cdot M^{\#(d)} \cdot \left(\frac12\right)^{\#(c) + \#(d)},
}
where the $\#$'s count the number of $c\+$'s, $d\+$'s, and $a\+b\+$ pairs in $\O$. The normalization factor $\mathcal N = 1/\sqrt{\qmat\Omega{\O\+\O}\Omega}$ is thus precisely what has been stated in eqs.~\eqref{c-norm}--\eqref{e-norm}.

\subsection{Perturbative calculation of the gap}

As an application of interest for studying the low-energy behavior of the system, we now show how the diagrammatic rules above can be used to calculate the first correction to the gap of the system.

The ground state only receives corrections from disconnected (bubble) diagrams. As usual, these do not affect the gap calculation; they merely renormalize the ground state energy which we may always shift to zero. (This is why we only need to compute connected diagrams, as per rule 2.) Thus, we take $E(\Omega) = E_0(\Omega) = 0$. The only corrections to the gap come from the corrections to $E(\mathcal C_2)$, the energy of the first excited state.

The state $\qvec{\mathcal C_2}$ has energy $E_0(\mathcal C_2) = 2m_{SU(N)}$, and so the unperturbed gap is
\bel{
  \Delta_0 = E_0(\mathcal C_2) - E_0(\Omega) = 2m_{SU(N)} = \frac{2m\xi}{1 + \xi} =  2\xi\sqrt{\lambda\pi} + O(\xi^2).
}
The first order corrections is trivially
\bel{
  E_1(\mathcal C_2) = \qmat{\mathcal C_2}V{\mathcal C_2} = 0.
}
The second order correction is non-trivial, and hence we resort to the Feynman rules developed in the previous section.

\begin{figure}[tb]
 \centering
 \includegraphics[scale = 0.8, clip=true,trim= 50pt 80pt 10pt 50pt]{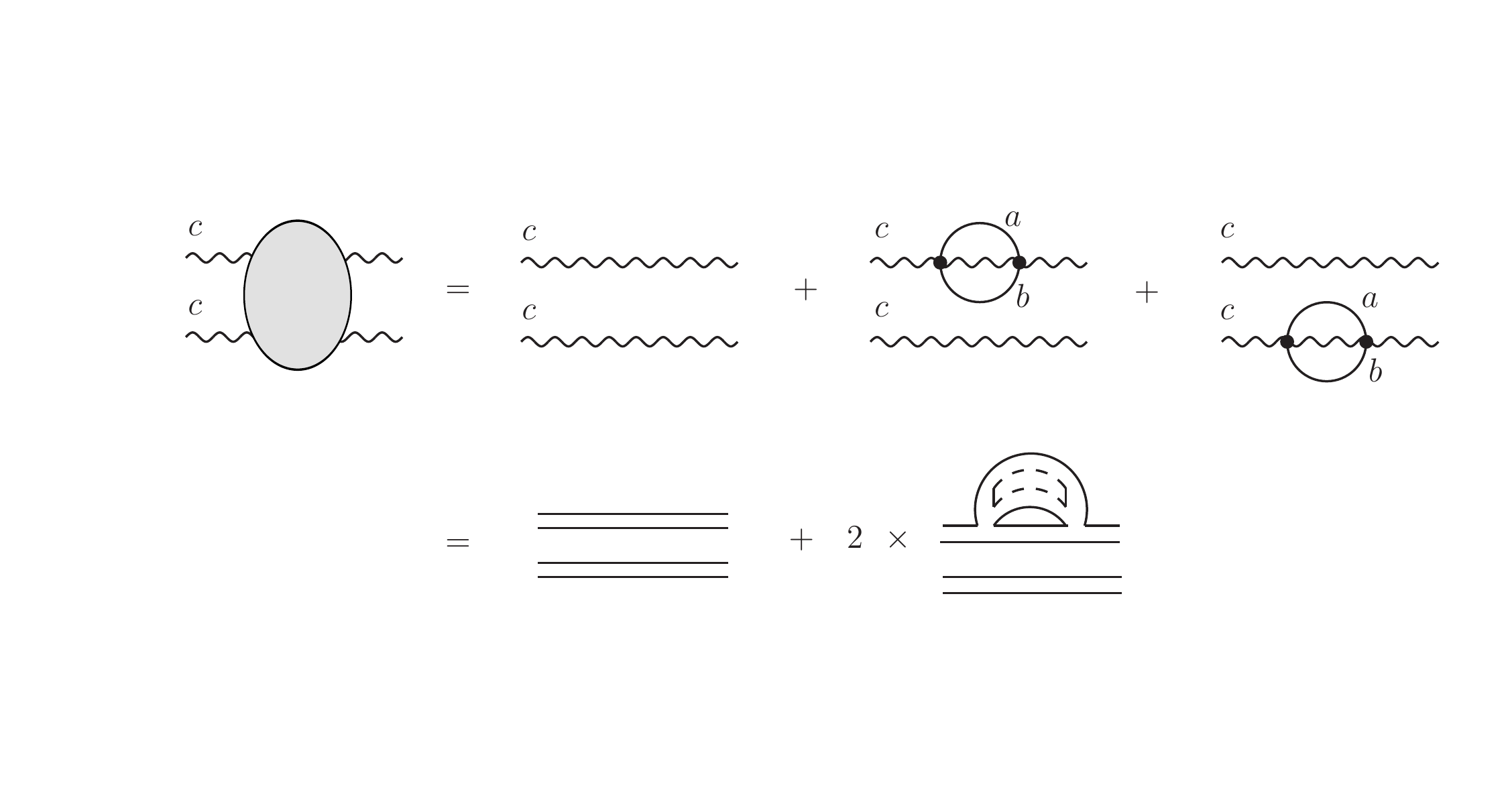}
 % feyndiag.pdf: 0x0 pixel, 0dpi, 0.00x0.00 cm, bb=
 \caption{\small The three diagrams that contribute to $E(\mathcal C_2)$ up to second order in perturbation theory, shown in single-line notation (above) and double-line notation (below). The first diagram is the zero-vertex (unperturbed) energy, and the other two are the only two-vertex diagrams that contribute in the 't Hooft limit.}
 \label{fig:feyndiag}
\end{figure}

We have already argued that $E_2(\mathcal C_2)$ scales like $\xi$, and is hence of the same order as the unperturbed energy. The diagrams contributing up to second order in perturbation theory are shown on Fig.~\ref{fig:feyndiag}. The zero-vertex term is just the unperturbed energy $E_0(\mathcal C_2)$. The second order term in perturbation theory (i.e.~the sum of two-vertex diagrams) is calculated following our diagrammatic rules.  Each of the two two-vertex diagrams has one $N$-loop, one $M$-loop, one gauge boson contraction, and two vertices. The intermediate state is in both cases $\qvec{\mathcal E_1^{2,0}}$, with $s = 1$ and $E_0(\mathcal E^{2,0}_1) - E_0(\mathcal C_2) = 2(m + m_{SU(N)})$, while the in- and out-states have $s = 2$. Hence, the two-vertex contribution, or the second order term in perturbation theory, is
\bel{
  E_2(\mathcal C_2) = - 2\cdot \frac12 \cdot \frac12 \cdot \left(\frac{g^2}{2m}\right)^2 \frac1{2(m + m_{SU(N)})} N M = -\frac{\xi m}{(1 + \xi)(1 + 2\xi)}.
}
We are working up to first order in $\xi$, and in this regime the second-order-corrected gap is
\bel{
  \Delta = \frac{2\xi m}{1+\xi} - \frac{\xi m}{(1 + \xi)(1 + 2\xi)} = \xi \sqrt{\lambda \pi} + O(\xi^2).
}
Higher order corrections can be found in a similarly straightforward way. Note that this answer reduces to the answer for CS-fundamental on a torus when $\xi = 1/N$.

%\section{Bifundamentals On A Genus, $g\ge 2$, Riemann Surface}

%In the previous section we saw that on a torus $(g=1)$ the energy of low lying states scales as $\sqrt\lambda \frac{M}{N}$, which has a nonzero value in the 't Hooft limit if we keep the ration $\frac{M}{N}$ fixed. This is in contrast to the fundamental matter where the energy scales like $\frac{\sqrt\lambda}{N}$ and goes to zero in the strict classical limit, $N=\infty$, in the bulk.

%Let us now place the theory on a genus, $g\ge 2$, Riemann surface, $\Sigma_g$. The number of states in the Hilbert space of pure Chern-Simons theory on $\Sigma_{g}$with gauge group $U(N)$ and level $k$, goes like $k^{(g-1)N^2}$ for large $k$ \cite{Witten:1988hf}. So for the bifundamental theory this will go like $k^{(g-1)(N^2 + M^2)}$. These states have exactly zero energy in the pure Chern-Simons theory. It will be very interesting to know the fate of these large number of exactly degenerate states once we add matter to it. If we add fundamental matter to it then the degeneracy is not lifted at least in the very weakly coupled regime. But for the bifundamental matter things could be different. We saw in the case of torus that adding bifundamental matter lifts the degeneracy of pure Chern-Simons states even in the classical limit in the bulk. The same thing could happen for higher genus surfaces and if this is the case then it will be fascinating. We leave this question for future study.

\section{Discussion}

In this paper we have studied the Chern-Simons theory coupled to bifundamental scalar fields. At two-loop order, in the 't Hooft limit, the theory has two lines of fixed points parametrized by the 't Hooft coupling. These lines exist for all values of the fixed ratio $M/N$. When this ratio is zero the theory goes over to the CS theory coupled to the fundamental matter which has a dual description in terms of a parity-violating version of the Vasiliev higher-spin gauge theory in $AdS_4$. When the ratio is small the dual gravitational theory should be some deformation of the Vasiliev theory, as has been conjectured in \cite{Chang:2012kt}. When this ratio approaches unity, the field theory is some kind of a non-supersymmetric version of the ABJM theory. The dual gravitational theory should be an Einstein gravity theory in the large 't Hooft coupling limit. It will be fascinating to better understand this bosonic theory in the bulk.

We have also studied the low-lying spectrum of this theory placed on a torus. CS theory coupled to fundamental matter has a set of low lying states whose energy goes like $\sqrt\lambda/N$ and so in the strict classical limit, $N=\infty$, they are exactly degenerate zero energy states. Our analysis shows that the bifundamental matter has no such states on the torus. The energy in this case goes like $\sqrt\lambda\ M/N$ for small value of $M/N$, and this gap stays nonzero even when $N=\infty$. This is encouraging and leads to the picture where one can think of the bifundamental theory as a regulator of the fundamental theory which regulates the singular low-energy states of the fundamental theory on a torus whose bulk dual is still mysterious.

Let us now place the theory on a genus $g\ge 2$ Riemann surface, $\Sigma_g$. The number of states in the Hilbert space of pure CS theory on $\Sigma_{g}$ with gauge group $U(N)$ and level $k$ goes like $k^{(g-1)N^2}$ for large $k$ \cite{Witten:1988hf}. So, for the bifundamental theory, this will go like $k^{(g-1)(N^2 + M^2)}$. These states have exactly zero energy in the pure CS theory. It will be very interesting to know the fate of these large number of exactly degenerate states once we add matter to it. If we add fundamental matter, then the degeneracy is not lifted at least in the very weakly coupled regime. However, for the bifundamental matter things could be different. We saw in the case of torus that adding bifundamental matter lifts the degeneracy of pure CS states even in the classical limit in the bulk. The same thing could happen for higher genus surfaces and if this is the case then it will be fascinating. We leave this question for future study.

We have left untouched many important things in this paper. For example, we have not provided any argument for the all-loop existence of the fixed line. One can also compute the anomalous dimensions of the operators, and it is known that the currents acquire non-zero anomalous dimensions in the bifundamental theory. It would be good to have an expression for that as a function of the 't Hooft coupling and the fixed ratio $M/N$. Another important thing is to compute the free energy of this theory. It will give us a wealth of information about the bulk gravity theory, in particular about black holes in the bulk. In passing we would like to mention that the bifundamental fermions coupled to CS gauge theory may be simpler in this respect, because by standard arguments the fermionic theory is a conformal field theory for all values of the 't Hooft coupling and the ratio $M/N$.

We have not touched upon the issue of duality \cite{Aharony:2012nh, Maldacena:2011jn, Maldacena:2012sf} in these bifundamental theories. For that one has to study the scalar and fermion theories in much more detail. We leave that for future research.

\section{Acknowledgement}
We would like to thank Shiraz Minwalla for suggesting this problem and Eric Perlmutter for comments on the previous version of the paper. We would also like to thank Guy Gur-Ari, Raghu Mahajan, Shiraz Minwalla, Xiao-Liang Qi, and Steve Shenker for valuable discussions.  We would also like to thank Steve Shenker for useful comments on the draft. The work of \DJ R was supported by the Stanford Institute for Theoretical Physics (SITP) and an NSF Graduate Research Fellowship. The research of SB is supported by NSF grant 0756174 and the SITP.

\appendix

\section{Feynman rules for CS-bifundamental theories} \label{sec:FeynRules}

The action of the CS-bifundamental theory is given by eq.~\eqref{def S}. The Feynman rules are easily obtained if the notation is decompactified and all indices are explicitly written. This gives
\algnl{
  S
  &= S_A + S_B + S\_{matter}, \qquad \trm{with}\\
  S_A
  &= \int \d^3 x \left(\frac i2 C(N) \epsilon_{\mu\nu\lambda} A_\mu^a \del_\nu A_\lambda^a - \frac i6 g C(N) \epsilon_{\mu\nu\lambda} f^{abc} A^a_\mu A_\nu^b A_\lambda^c \right),\\
  S_B
  &= \int \d^3 x \left(-\frac i2 C(M) \epsilon_{\mu\nu\lambda} B_\mu^\alpha \del_\nu B_\lambda^\alpha + \frac i6 g C(M) \epsilon_{\mu\nu\lambda} f^{\alpha\beta\gamma} B^\alpha_\mu B_\nu^\beta B_\lambda^\gamma \right),\\ \notag
  S\_{matter}
  &= \int \d^3 x \bigg(\del \phi\+_{ii'} \del \phi_{ii'} + 2ig \ \phi_{ii'}\+ A^a_\mu T^a_{ij} \del_\mu \phi_{ji'} + 2ig\ \phi_{ii'} B^\alpha_\mu T^\alpha_{i'j'}  \del_\mu \phi\+_{ij'} + \\ \notag
  & + g^2 \phi\+_{ii'} A^a_\mu T^a_{ij} A^b_\mu T^b_{jk} \phi_{ki'} + g^2 \phi_{ii'} B^\alpha_\mu T^\alpha_{i'j'} B^\beta_\mu T^\beta_{j'k'} \phi\+_{ik'} - 2g^2 \phi\+_{ii'} A_\mu^a T^a_{ij} \phi_{jj'} B_\mu^\alpha T^\alpha_{j'i'} + \\
  &+ \frac{g_1}3 \phi\+_{ii'} \phi_{ij'} \phi\+_{jj'} \phi_{jk'} \phi\+_{kk'} \phi_{ki'} + \frac{g_2}2 \phi\+_{ii'} \phi_{ij'} \phi\+_{jj'} \phi_{ji'} \phi\+_{kk'} \phi_{kk'} + \frac{g_3}6 \phi\+_{ii'} \phi_{ii'} \phi\+_{jj'} \phi_{jj'} \phi\+_{kk'} \phi_{kk'}\bigg).
}
As elsewhere in the text, $a$, $b$, $c$, etc.~run over generators of $SU(N)$, $\alpha$, $\beta$, $\gamma$, etc.~run over generators of $SU(M)$. The mid-alphabet Roman indices $i$, $j$, $k$, etc.~run over the components of $SU(N)$ vectors in the fundamental representation, and the corresponding primed indices run over components of $SU(M)$ fundamentals. Everything is done in Landau gauge, $\del_\mu A_\mu = 0$. We have defined the CS coupling
\bel{
  g^2 \equiv \frac{4\pi}k,
}
rescaled all the gauge fields by $g$, and defined
\bel{
  \Tr_N(T^a T^b) = C(N) \delta^{ab},\quad \Tr_N\left([T^a, T^b] T^c \right) = i C(N) f^{abc}.
}
In this paper, consistently with the choice of the CS action in Section \ref{sec:feyn2loops} and following \cite{Aharony:2011jz}, we choose
\bel{
  C(N) = 1 \quad\trm{and}\quad C_2(N) = N + O\left(\frac1N\right),
}
where the latter equality for $C_2(N) \equiv T^a T^a$ comes as a special case of the choice
\bel{
  T^a_{ij} T^a_{kl} = \delta_{il} \delta_{jk} + O\left(\frac1N\right).
}
The Feynman rules can now be read off. The propagators are:
\algnl{
  \avg{A_\mu^a(p) A_\nu^b(q)}_0
  &= \delta^{ab} \epsilon_{\mu\lambda\nu} \frac{p^\lambda}{p^2}(2\pi)^3 \delta^3(p - q),\\
  \avg{B_\mu^\alpha(p) B_\nu^\beta(q)}_0
  &= - \delta^{\alpha\beta} \epsilon_{\mu\lambda\nu} \frac{p^\lambda}{p^2}(2\pi)^3 \delta^3(p - q), \\
  \avg{\phi\+_{ii'}(p) \phi_{jj'}(q)}_0
  &= \delta_{ij} \delta_{i'j'} \frac1{p^2} (2\pi)^3 \delta^3(p - q). }
Note that the gluon propagators must be assigned a direction. Taking into account the overall minus sign from the Boltzmann factor $e^{-S}$, the vertices are found to be:
\algnl{\notag
  &\avg{\phi_{ii'}\+ (p_1) \phi_{jj'}(p_2) \phi_{kk'}\+ (p_3) \phi_{ll'} (p_4) \phi_{mm'}\+ (p_5) \phi_{nn'}(p_6)}_0 \\ \notag
  &\quad = - g_1 \left(\delta_{i'j'} \delta_{jk} \delta_{k'l'} \delta_{lm} \delta_{m'n'} \delta_{ni} + \trm{11\ other\ permutations} \right)(2\pi)^3 \delta^3\left(\sum p_i\right) - \\ \notag
  &\qquad - g_2 \left(\delta_{i'j'} \delta_{jk} \delta_{k'l'} \delta_{li} \delta_{m'n'} \delta_{nm} + \trm{17\ other\ permutations} \right)(2\pi)^3 \delta^3\left(\sum p_i\right) - \\
  &\qquad - g_3 \left(\delta_{i'j'} \delta_{ji} \delta_{k'l'} \delta_{lk} \delta_{m'n'} \delta_{nm} + \trm{5\ other \ permutations} \right)(2\pi)^3 \delta^3\left(\sum p_i\right),\\
  &\avg{\phi_{ii'}\+ (p_1) A_\mu^a (p_2) \phi_{jj'}(p_3)}_0
  = -2 g\ p_3{}_\mu\ \delta_{i'j'} T^a_{ij}\ (2\pi)^3 \delta^3\left(\sum p_i\right),\\
  &\avg{\phi_{ii'} (p_1) B_\mu^\alpha (p_2) \phi_{jj'}\+(p_3)}_0
  = -2g\ p_3{}_\mu\ \delta_{ij} T^\alpha_{i'j'}\ (2\pi)^3 \delta^3\left(\sum p_i\right),\\
  &\avg{\phi_{ii'}\+ (p_1) A_\mu^a (p_2) A_\nu^b(p_3) \phi_{jj'}(p_4)}_0
  = -g^2\ \delta_{i'j'}\ \delta_{\mu\nu}\ \left(T^a_{ik} T^b_{kj} + T^b_{ik} T^a_{kj}\right)\ (2\pi)^3 \delta^3\left(\sum p_i\right),\\
  &\avg{\phi_{ii'} (p_1) B_\mu^\alpha (p_2) B_\nu^\beta(p_3) \phi_{jj'}\+(p_4)}_0
  = -g^2\ \delta_{ij}\ \delta_{\mu\nu}\ \left(T^\alpha_{i'k'} T^\beta_{k'j'} + T^\beta_{i'k'} T^\alpha_{k'j'}\right)\ (2\pi)^3 \delta^3\left(\sum p_i\right),\\
  &\avg{\phi_{ii'}\+ (p_1) A_\mu^a (p_2) \phi_{jj'}(p_3) B_\nu^\alpha(p_4)}_0
  = 2g^2\ \delta_{\mu\nu}\ T^a_{ij} T^\alpha_{j'i'}\ (2\pi)^3 \delta^3\left(\sum p_i\right),\\
  &\avg{A_\mu^a(p_1) A_\nu^b(p_2) A_\lambda^c (p_3)}_0
  = ig\ \epsilon_{\mu\nu\lambda} f^{abc} \ (2\pi)^3 \delta^3\left(\sum p_i\right),\\
  &\avg{B_\mu^\alpha(p_1) B_\nu^\beta(p_2) B_\lambda^\gamma (p_3)}_0 = -ig\ \epsilon_{\mu\nu\lambda} f^{\alpha\beta\gamma} \ (2\pi)^3 \delta^3\left(\sum p_i\right).
}
These can be depicted using a variant of the usual double-line notation with two types of lines. We introduce these diagrams in Section \ref{sec:torus}, Fig.~\ref{fig:doublelines}. This notation has the usual advantage of allowing one to quickly estimate which diagrams dominate in the 't Hooft limit. Note that we do not include any Feynman rules for ghosts because these do not appear in any of the diagrams relevant for our purposes.


\begin{thebibliography}{99}


  %\cite{Vasiliev:1992av}
\bibitem{Vasiliev:1992av}
  M.~A.~Vasiliev,
  ``More on equations of motion for interacting massless fields of all spins in (3+1)-dimensions,''
  Phys.\ Lett.\ B {\bf 285}, 225 (1992).
  %%CITATION = PHLTA,B285,225;%%
  %170 citations counted in INSPIRE as of 12 Jul 2013


  %\cite{Vasiliev:1995dn}
\bibitem{Vasiliev:1995dn}
  M.~A.~Vasiliev,
  ``Higher spin gauge theories in four-dimensions, three-dimensions, and two-dimensions,''
  Int.\ J.\ Mod.\ Phys.\ D {\bf 5}, 763 (1996)
  [hep-th/9611024].
  %%CITATION = HEP-TH/9611024;%%
  %192 citations counted in INSPIRE as of 12 Jul 2013


  %\cite{Vasiliev:1999ba}
\bibitem{Vasiliev:1999ba}
  M.~A.~Vasiliev,
  ``Higher spin gauge theories: Star product and AdS space,''
  In *Shifman, M.A. (ed.): The many faces of the superworld* 533-610
  [hep-th/9910096].
  %%CITATION = HEP-TH/9910096;%%
  %230 citations counted in INSPIRE as of 12 Jul 2013


  %\cite{Vasiliev:2003ev}
\bibitem{Vasiliev:2003ev}
  M.~A.~Vasiliev,
  ``Nonlinear equations for symmetric massless higher spin fields in (A)dS(d),''
  Phys.\ Lett.\ B {\bf 567}, 139 (2003)
  [hep-th/0304049].
  %%CITATION = HEP-TH/0304049;%%
  %246 citations counted in INSPIRE as of 12 Jul 2013


%\cite{Sezgin:2002ru}
\bibitem{Sezgin:2002ru}
  E.~Sezgin and P.~Sundell,
  ``Analysis of higher spin field equations in four-dimensions,''
  JHEP {\bf 0207}, 055 (2002)
  [hep-th/0205132].
  %%CITATION = HEP-TH/0205132;%%
  %66 citations counted in INSPIRE as of 12 Jul 2013


   %\cite{Giombi:2012ms}
\bibitem{Giombi:2012ms}
  S.~Giombi and X.~Yin,
  ``The Higher Spin/Vector Model Duality,''
  J.\ Phys.\ A {\bf 46}, 214003 (2013)
  [arXiv:1208.4036 [hep-th]].
  %%CITATION = ARXIV:1208.4036;%%
  %36 citations counted in INSPIRE as of 12 Jul 2013


   %\cite{Gaberdiel:2010pz}
\bibitem{Gaberdiel:2010pz}
  M.~R.~Gaberdiel and R.~Gopakumar,
  ``An $AdS_3$ Dual for Minimal Model CFTs,''
  Phys.\ Rev.\ D {\bf 83}, 066007 (2011)
  [arXiv:1011.2986 [hep-th]].
  %%CITATION = ARXIV:1011.2986;%%
  %130 citations counted in INSPIRE as of 12 Jul 2013


  %\cite{Gaberdiel:2013vva}
\bibitem{Gaberdiel:2013vva}
  M.~R.~Gaberdiel and R.~Gopakumar,
  ``Large $\mathcal{N}=4$ Holography,''
  arXiv:1305.4181 [hep-th].
  %%CITATION = ARXIV:1305.4181;%%
  %3 citations counted in INSPIRE as of 12 Jul 2013

%\cite{Chang:2011mz}
\bibitem{Chang:2011mz}
  C.-M.~Chang and X.~Yin,
  ``Higher Spin Gravity with Matter in $AdS_3$ and Its CFT Dual,''
  JHEP {\bf 1210}, 024 (2012)
  [arXiv:1106.2580 [hep-th]].
  %%CITATION = ARXIV:1106.2580;%%
  %68 citations counted in INSPIRE as of 12 Jul 2013


  %\cite{Chang:2013izp}
\bibitem{Chang:2013izp}
  C.-M.~Chang and X.~Yin,
  ``A semi-local holographic minimal model,''
  arXiv:1302.4420 [hep-th].
  %%CITATION = ARXIV:1302.4420;%%
  %8 citations counted in INSPIRE as of 12 Jul 2013


  %\cite{Anninos:2011ui}
\bibitem{Anninos:2011ui}
  D.~Anninos, T.~Hartman and A.~Strominger,
  ``Higher Spin Realization of the dS/CFT Correspondence,''
  arXiv:1108.5735 [hep-th].
  %%CITATION = ARXIV:1108.5735;%%
  %56 citations counted in INSPIRE as of 12 Jul 2013


  %\cite{Ng:2012xp}
\bibitem{Ng:2012xp}
  G.~S.~Ng and A.~Strominger,
  ``State/Operator Correspondence in Higher-Spin dS/CFT,''
  Class.\ Quant.\ Grav.\  {\bf 30}, 104002 (2013)
  [arXiv:1204.1057 [hep-th]].
  %%CITATION = ARXIV:1204.1057;%%
  %19 citations counted in INSPIRE as of 12 Jul 2013


  %\cite{Anninos:2012ft}
\bibitem{Anninos:2012ft}
  D.~Anninos, F.~Denef and D.~Harlow,
  ``The Wave Function of Vasiliev's Universe - A Few Slices Thereof,''
  arXiv:1207.5517 [hep-th].
  %%CITATION = ARXIV:1207.5517;%%
  %12 citations counted in INSPIRE as of 12 Jul 2013


  %\cite{Anninos:2013rza}
\bibitem{Anninos:2013rza}
  D.~Anninos, F.~Denef, G.~Konstantinidis and E.~Shaghoulian,
  ``Higher Spin de Sitter Holography from Functional Determinants,''
  arXiv:1305.6321 [hep-th].
  %%CITATION = ARXIV:1305.6321;%%
  %3 citations counted in INSPIRE as of 12 Jul 2013


  %\cite{Banerjee:2013mca}
\bibitem{Banerjee:2013mca}
  S.~Banerjee, A.~Belin, S.~Hellerman, A.~Lepage-Jutier, A.~Maloney, \DJ.~Radi\v cevi\'c and S.~Shenker,
  ``Topology of Future Infinity in dS/CFT,''
  arXiv:1306.6629 [hep-th].
  %%CITATION = ARXIV:1306.6629;%%


  %\cite{Das:2012dt}
\bibitem{Das:2012dt}
  D.~Das, S.~R.~Das, A.~Jevicki and Q.~Ye,
  ``Bi-local Construction of Sp(2N)/dS Higher Spin Correspondence,''
  JHEP {\bf 1301}, 107 (2013)
  [arXiv:1205.5776 [hep-th]].
  %%CITATION = ARXIV:1205.5776;%%
  %11 citations counted in INSPIRE as of 12 Jul 2013

  %\cite{Klebanov:2002ja}
\bibitem{Klebanov:2002ja}
  I.~R.~Klebanov and A.~M.~Polyakov,
  ``AdS dual of the critical O(N) vector model,''
  Phys.\ Lett.\ B {\bf 550}, 213 (2002)
  [hep-th/0210114].
  %%CITATION = HEP-TH/0210114;%%

%\cite{Sezgin:2002rt}
\bibitem{Sezgin:2002rt}
  E.~Sezgin and P.~Sundell,
  ``Massless higher spins and holography,''
  Nucl.\ Phys.\ B {\bf 644}, 303 (2002)
  [Erratum-ibid.\ B {\bf 660}, 403 (2003)]
  [hep-th/0205131].
  %%CITATION = HEP-TH/0205131;%%
  %238 citations counted in INSPIRE as of 12 Jul 2013

  %\cite{Petkou:2003zz}
\bibitem{Petkou:2003zz}
  A.~C.~Petkou,
  ``Evaluating the AdS dual of the critical O(N) vector model,''
  JHEP {\bf 0303}, 049 (2003)
  [hep-th/0302063].
  %%CITATION = HEP-TH/0302063;%%
  %50 citations counted in INSPIRE as of 12 Jul 2013

%\cite{Sezgin:2003pt}
\bibitem{Sezgin:2003pt}
  E.~Sezgin and P.~Sundell,
  ``Holography in 4D (super) higher spin theories and a test via cubic scalar couplings,''
  JHEP {\bf 0507}, 044 (2005)
  [hep-th/0305040].
  %%CITATION = HEP-TH/0305040;%%
  %100 citations counted in INSPIRE as of 12 Jul 2013

  %\cite{Girardello:2002pp}
\bibitem{Girardello:2002pp}
  L.~Girardello, M.~Porrati and A.~Zaffaroni,
  ``3-D interacting CFTs and generalized Higgs phenomenon in higher spin theories on AdS,''
  Phys.\ Lett.\ B {\bf 561}, 289 (2003)
  [hep-th/0212181].
  %%CITATION = HEP-TH/0212181;%%
  %49 citations counted in INSPIRE as of 12 Jul 2013


%\cite{Giombi:2009wh}
\bibitem{Giombi:2009wh}
  S.~Giombi and X.~Yin,
  ``Higher Spin Gauge Theory and Holography: The Three-Point Functions,''
  JHEP {\bf 1009}, 115 (2010)
  [arXiv:0912.3462 [hep-th]].
  %%CITATION = ARXIV:0912.3462;%%

  %\cite{Giombi:2010vg}
\bibitem{Giombi:2010vg}
  S.~Giombi and X.~Yin,
  "Higher Spins in AdS and Twistorial Holography,''
  JHEP {\bf 1104}, 086 (2011)
  [arXiv:1004.3736 [hep-th]].
  %%CITATION = ARXIV:1004.3736;%%
  %102 citations counted in INSPIRE as of 12 Jul 2013


  %\cite{Giombi:2011ya}
\bibitem{Giombi:2011ya}
  S.~Giombi and X.~Yin,
  ``On Higher Spin Gauge Theory and the Critical O(N) Model,''
  Phys.\ Rev.\ D {\bf 85}, 086005 (2012)
  [arXiv:1105.4011 [hep-th]].
  %%CITATION = ARXIV:1105.4011;%%
  %49 citations counted in INSPIRE as of 12 Jul 2013

  %\cite{Chang:2012kt}
\bibitem{Chang:2012kt}
  C.-M.~Chang, S.~Minwalla, T.~Sharma and X.~Yin,
  ``ABJ Triality: from Higher Spin Fields to Strings,''
  arXiv:1207.4485 [hep-th].
  %%CITATION = ARXIV:1207.4485;%%
  %32 citations counted in INSPIRE as of 28 Apr 2013



  %\cite{Das:2003vw}
\bibitem{Das:2003vw}
  S.~R.~Das and A.~Jevicki,
  ``Large N collective fields and holography,''
  Phys.\ Rev.\ D {\bf 68}, 044011 (2003)
  [hep-th/0304093].
  %%CITATION = HEP-TH/0304093;%%
  %40 citations counted in INSPIRE as of 12 Jul 2013



  %\cite{Koch:2010cy}
\bibitem{Koch:2010cy}
  R.~d.~M.~Koch, A.~Jevicki, K.~Jin and J.~P.~Rodrigues,
  ``$AdS_4/CFT_3$ Construction from Collective Fields,''
  Phys.\ Rev.\ D {\bf 83}, 025006 (2011)
  [arXiv:1008.0633 [hep-th]].
  %%CITATION = ARXIV:1008.0633;%%
  %62 citations counted in INSPIRE as of 12 Jul 2013

%\cite{Douglas:2010rc}
\bibitem{Douglas:2010rc}
  M.~R.~Douglas, L.~Mazzucato and S.~S.~Razamat,
  ``Holographic dual of free field theory,''
  Phys.\ Rev.\ D {\bf 83}, 071701 (2011)
  [arXiv:1011.4926 [hep-th]].
  %%CITATION = ARXIV:1011.4926;%%
  %69 citations counted in INSPIRE as of 12 Jul 2013


%\cite{Jevicki:2011ss}
\bibitem{Jevicki:2011ss}
  A.~Jevicki, K.~Jin and Q.~Ye,
  ``Collective Dipole Model of AdS/CFT and Higher Spin Gravity,''
  J.\ Phys.\ A {\bf 44}, 465402 (2011)
  [arXiv:1106.3983 [hep-th]].
  %%CITATION = ARXIV:1106.3983;%%
  %28 citations counted in INSPIRE as of 12 Jul 2013




%\cite{Giombi:2011kc}
\bibitem{Giombi:2011kc}
  S.~Giombi, S.~Minwalla, S.~Prakash, S.~P.~Trivedi, S.~R.~Wadia and X.~Yin,
  ``Chern-Simons Theory with Vector Fermion Matter,''
  Eur.\ Phys.\ J.\ C {\bf 72}, 2112 (2012)
  [arXiv:1110.4386 [hep-th]].
  %%CITATION = ARXIV:1110.4386;%%
  %45 citations counted in INSPIRE as of 28 Apr 2013


%\cite{Aharony:2011jz}
\bibitem{Aharony:2011jz}
  O.~Aharony, G.~Gur-Ari and R.~Yacoby,
  ``$d=3$ Bosonic Vector Models Coupled to Chern-Simons Gauge Theories,''
  JHEP {\bf 1203}, 037 (2012)
  [arXiv:1110.4382 [hep-th]].
  %%CITATION = ARXIV:1110.4382;%%
  %39 citations counted in INSPIRE as of 08 Jun 2013


   %\cite{Shenker:2011zf}
\bibitem{Shenker:2011zf}
  S.~H.~Shenker and X.~Yin,
  ``Vector Models in the Singlet Sector at Finite Temperature,''
  arXiv:1109.3519 [hep-th].
  %%CITATION = ARXIV:1109.3519;%%
  %30 citations counted in INSPIRE as of 12 Jul 2013

  %\cite{Banerjee:2012gh}
\bibitem{Banerjee:2012gh}
  S.~Banerjee, S.~Hellerman, J.~Maltz and S.~H.~Shenker,
  ``Light States in Chern-Simons Theory Coupled to Fundamental Matter,''
  arXiv:1207.4195 [hep-th].
  %%CITATION = ARXIV:1207.4195;%%


%\cite{Witten:1988hf}
\bibitem{Witten:1988hf}
  E.~Witten,
  ``Quantum Field Theory and the Jones Polynomial,''
  Commun.\ Math.\ Phys.\  {\bf 121}, 351 (1989).
  %%CITATION = CMPHA,121,351;%%
  %1936 citations counted in INSPIRE as of 12 Jul 2013

  %\cite{Elitzur:1989nr}
\bibitem{Elitzur:1989nr}
  S.~Elitzur, G.~W.~Moore, A.~Schwimmer and N.~Seiberg,
  ``Remarks on the Canonical Quantization of the Chern-Simons-Witten Theory,''
  Nucl.\ Phys.\ B {\bf 326}, 108 (1989).
  %%CITATION = NUPHA,B326,108;%%
  %398 citations counted in INSPIRE as of 12 Jul 2013


  %\cite{Maldacena:2011jn}
\bibitem{Maldacena:2011jn}
  J.~Maldacena and A.~Zhiboedov,
  ``Constraining Conformal Field Theories with A Higher Spin Symmetry,''
  J.\ Phys.\ A {\bf 46}, 214011 (2013)
  [arXiv:1112.1016 [hep-th]].
  %%CITATION = ARXIV:1112.1016;%%
  %63 citations counted in INSPIRE as of 12 Jul 2013


  %\cite{Maldacena:2012sf}
\bibitem{Maldacena:2012sf}
  J.~Maldacena and A.~Zhiboedov,
  ``Constraining conformal field theories with a slightly broken higher spin symmetry,''
  Class.\ Quant.\ Grav.\  {\bf 30}, 104003 (2013)
  [arXiv:1204.3882 [hep-th]].
  %%CITATION = ARXIV:1204.3882;%%
  %50 citations counted in INSPIRE as of 12 Jul 2013

  %\cite{Aharony:2012nh}
\bibitem{Aharony:2012nh}
  O.~Aharony, G.~Gur-Ari and R.~Yacoby,
  ``Correlation Functions of Large N Chern-Simons-Matter Theories and Bosonization in Three Dimensions,''
  JHEP {\bf 1212}, 028 (2012)
  [arXiv:1207.4593 [hep-th]].
  %%CITATION = ARXIV:1207.4593;%%
  %20 citations counted in INSPIRE as of 12 Jul 2013


  %\cite{Radicevic:2012in}
\bibitem{Radicevic:2012in}
  \DJ.~Radi\v cevi\'c,
  ``Singlet Vector Models on Lens Spaces,''
  arXiv:1210.0255 [hep-th].
  %%CITATION = ARXIV:1210.0255;%%
  %2 citations counted in INSPIRE as of 12 Jul 2013

 %\cite{Chen:1992ee}
\bibitem{Chen:1992ee}
  W.~Chen, G.~W.~Semenoff and Y.~-S.~Wu,
  ``Two loop analysis of nonAbelian Chern-Simons theory,''
  Phys.\ Rev.\ D {\bf 46}, 5521 (1992)
  [hep-th/9209005].
  %%CITATION = HEP-TH/9209005;%%
  %93 citations counted in INSPIRE as of 12 Jul 2013




  %\cite{Aharony:2012ns}
\bibitem{Aharony:2012ns}
  O.~Aharony, S.~Giombi, G.~Gur-Ari, J.~Maldacena and R.~Yacoby,
  ``The Thermal Free Energy in Large N Chern-Simons-Matter Theories,''
  JHEP {\bf 1303}, 121 (2013)
  [arXiv:1211.4843 [hep-th]].
  %%CITATION = ARXIV:1211.4843;%%
  %9 citations counted in INSPIRE as of 12 Jul 2013


  %\cite{Jain:2013py}
\bibitem{Jain:2013py}
  S.~Jain, S.~Minwalla, T.~Sharma, T.~Takimi, S.~R.~Wadia and S.~Yokoyama,
  ``Phases of large $N$ vector Chern-Simons theories on $S^2 \times S^1$,''
  arXiv:1301.6169 [hep-th].
  %%CITATION = ARXIV:1301.6169;%%
  %7 citations counted in INSPIRE as of 12 Jul 2013


  %\cite{Jain:2013gza}
\bibitem{Jain:2013gza}
  S.~Jain, S.~Minwalla and S.~Yokoyama,
  ``Chern Simons duality with a fundamental boson and fermion,''
  arXiv:1305.7235 [hep-th].
  %%CITATION = ARXIV:1305.7235;%%


  %\cite{Takimi:2013zca}
\bibitem{Takimi:2013zca}
  T.~Takimi,
  ``Duality and Higher Temperature Phases of Large $N$ Chern-Simons Matter Theories on $S^2 \times S^1$,''
  arXiv:1304.3725 [hep-th].
  %%CITATION = ARXIV:1304.3725;%%
  %3 citations counted in INSPIRE as of 12 Jul 2013


  %\cite{Jain:2012qi}
\bibitem{Jain:2012qi}
  S.~Jain, S.~P.~Trivedi, S.~R.~Wadia and S.~Yokoyama,
  ``Supersymmetric Chern-Simons Theories with Vector Matter,''
  JHEP {\bf 1210}, 194 (2012)
  [arXiv:1207.4750 [hep-th]].
  %%CITATION = ARXIV:1207.4750;%%
  %10 citations counted in INSPIRE as of 12 Jul 2013

  %\cite{Yokoyama:2012fa}
\bibitem{Yokoyama:2012fa}
  S.~Yokoyama,
  ``Chern-Simons-Fermion Vector Model with Chemical Potential,''
  JHEP {\bf 1301}, 052 (2013)
  [arXiv:1210.4109 [hep-th]].
  %%CITATION = ARXIV:1210.4109;%%
  %5 citations counted in INSPIRE as of 12 Jul 2013


  %\cite{GurAri:2012is}
\bibitem{GurAri:2012is}
  G.~Gur-Ari and R.~Yacoby,
  ``Correlators of Large N Fermionic Chern-Simons Vector Models,''
  JHEP {\bf 1302}, 150 (2013)
  [arXiv:1211.1866 [hep-th]].
  %%CITATION = ARXIV:1211.1866;%%
  %5 citations counted in INSPIRE as of 12 Jul 2013


  %\cite{Giombi:2013yva}
\bibitem{Giombi:2013yva}
  S.~Giombi, I.~R.~Klebanov, S.~S.~Pufu, B.~R.~Safdi and G.~Tarnopolsky,
  ``AdS Description of Induced Higher-Spin Gauge Theory,''
  arXiv:1306.5242 [hep-th].
  %%CITATION = ARXIV:1306.5242;%%


\end{thebibliography}
\end{document}